\documentclass[final,5p,times,twocolumn]{elsarticle}

\usepackage{amssymb}
\usepackage{amsmath}
\usepackage[hidelinks]{hyperref}
\usepackage{listings}
\usepackage{xcolor}   

\journal{Computer Communications}

\begin{document}

\begin{frontmatter}



\title{EAP-FIDO: A Novel EAP Method for Using FIDO2 Credentials for Network Authentication}


\author[rnasa,citic]{Martiño Rivera-Dourado} 

\author[unipi]{Christos Xenakis}

\author[rnasa,citic,inibic]{Alejandro Pazos}

\author[rnasa,citic]{Jose Vázquez-Naya}

\affiliation[rnasa]{organization={RNASA-IMEDIR, Universidade da Coruña},
            addressline={Facultade de Informática, Campus de Elviña}, 
            city={A Coruña},
            postcode={15071}, 
            state={Galicia},
            country={Spain}}

\affiliation[unipi]{organization={Department of Digital Systems},
	addressline={University of Piraeus}, 
	city={Piraeus},
	postcode={18532}, 
	state={Attica},
	country={Greece}}

\affiliation[citic]{organization={Centro de Investigación CITIC, Universidade da Coruña},
	addressline={Campus de Elviña}, 
	city={A Coruña},
	postcode={15071}, 
	state={Galicia},
	country={Spain}}

\affiliation[inibic]{organization={Biomedical Research Institute of A Coruña (INIBIC)},
	addressline={University Hospical Complex (CHUAC)}, 
	city={A Coruña},
	postcode={15006}, 
	state={Galicia},
	country={Spain}}

\begin{abstract}
The adoption of FIDO2 authentication by major tech companies in web applications has grown significantly in recent years. However, we argue FIDO2 has broader potential applications. In this paper, we introduce EAP-FIDO, a novel Extensible Authentication Protocol (EAP) method for use in IEEE 802.1X-protected networks. This allows organisations with WPA2/3-Enterprise wireless networks or MACSec-enabled wired networks to leverage FIDO2's passwordless authentication in compliance with existing standards. Additionally, we provide a comprehensive security and performance analysis to support the feasibility of this approach.
\end{abstract}

%


\end{frontmatter}



\section{Introduction}
\label{sec:intro}

Passwordless authentication is becoming a new approach in nowadays systems. The FIDO2 project is conducted by an alliance of some of the most important tech companies, aiming to replace the password as an authentication method with a new technology: Fast IDentity Online (FIDO) \cite{angelogianni_how_2024}.

Nowadays, the second generation of this technology (FIDO2) is being applied in some of the most common web applications, and being adopted by the main operating systems. Although there are still several obstacles for its complete adoption, FIDO2 will play a role in the future of authentication systems in many corporations \cite{lassak_why_2024}.

From credential stealing via phishing to keylogging, passwords are still vulnerable to multiple attacks \cite{veroni_large-scale_2022} \cite{wahab_investigating_2024}. Since 2014, the FIDO Alliance worked towards designing, developing and standardising the new passwordless future.

FIDO2 is a challenge-response protocol based on public-key cryptography, where an authenticator  holds the FIDO2 credential to sign the challenge received from the server. This authenticator may be an external USB/NFC/BLE security key or a platform-based authenticator that stores the credential within the operating system \cite{bradley_web_2021}. 

FIDO2 standards started to be adopted from the publication of the WebAuthn standard by the W3C in 2019 \cite{brand_web_2019}, which enables web applications to use the FIDO2 protocol for user authentication. However, there are few works that have applied FIDO2 to other scenarios, like network authentication \cite{rivera-dourado_novel_2024} \cite{huseynov_passwordless_2022}.

IEEE 802.1X \cite{noauthor_ieee_2020} is one of the most common approaches for Port-Based Network Access Control (PBNAC) nowadays. The Extensible Authentication Protocol (EAP) is created as an authentication framework within IEEE 802.1X \cite{vollbrecht_extensible_2004}, allowing different methods for user authentication. Nowadays, most EAP methods are still based on password authentication, like EAP-MD5, LEAP or some tunnelled methods based on PEAP or EAP-TTLS, like MSCHAPv2 or PAP.

EAP on 802.1X is used for authentication mostly in corporate networks. Thanks to the authentication framework, the EAP methods can be used on wireless networks with WPA2/3-Enterprise and on cabled networks with MACSec.

\subsection{Contribution}
In this paper, we propose EAP-FIDO: a novel user authentication protocol with FIDO2 credentials based on the EAP framework. Here, we provide (1) the protocol design, (2) a prototype implementation, (3) a performance analysis and, finally, (4) a security analysis.

EAP-FIDO is designed for authentication in corporate networks, like the Eduroam federated Wi-Fi network (GÉANT). Thanks to our design approach based in the EAP framework, EAP-FIDO works out-of-the-box in these networks by adding a compatible client and server, remaining transparent for the Access Points (APs). 

For instance, EAP-FIDO can be deployed in WPA2/3-Enterprise-enabled Wi-Fi networks, such as Eduroam, allowing registered users to authenticate using a passkey, also known as a FIDO2 credential. This credential can be securely stored on devices like a Yubikey or an Android 14+ smartphone, providing a convenient, passwordless authentication method.

There are several benefits of EAP-FIDO, beginning with security advantages (see section 6). However, these are not the only benefits of EAP-FIDO. By using this protocol, users take advantage of a usable passwordless authentication flow, which is present in web applications they are used to. In fact, organisations can use the same user FIDO2 credentials registered for web Single Sign On (SSO) with WebAuthn to provide network authentication with EAP-FIDO, achieving a coherent user experience.

Finally, our contribution follows the methodology of recent similar works (see section 8) based on the EAP framework, like Mortágua et al. \cite{mortagua_enhancing_2024}. In this paper, we use EAP to integrate the last FIDO2 standard: FIDO CTAP2 (Client To Authenticator Protocol).

\subsection{Document structure}
This paper is organized as follows: Section~\ref{sec:intro} outlines the research objectives and summarizes the primary contributions of this work. Section~\ref{sec:background} provides a comprehensive discussion of the related background. The design, architecture, and operational details of the EAP-FIDO protocol are presented in Section~\ref{sec:eap-fido}. Section~\ref{sec:prototype-implementation} describes the prototype implementation in detail, with its performance evaluated under real-world conditions in Section~\ref{sec:performance-analysis-and-results}. A detailed security analysis is conducted in Section~\ref{sec:security-analysis}, followed by a discussion of potential use cases of EAP-FIDO in Section~\ref{sec:use-cases}. Related work is examined in Section~\ref{sec:related-work}, and conclusions along with directions for future research are outlined in Section~\ref{sec:conclusions-and-future-work}.

\section{Background}
\label{sec:background}

EAP-FIDO is intended for use in enterprise networks within the Extensible Authentication Protocol (EAP) framework. This section provides an overview of the relevant background, including enterprise network authentication and the EAP framework. 

\subsection{Enterprise network authentication}
Within any network, substantial amounts of digital data are stored and exchanged between hosts, making both in-transit and at-rest data potentially accessible to devices connected to the network. Being inside a corporate network allows attackers to exploit vulnerabilities for attacks like eavesdropping on sensitive data or launching Denial of Service (DoS) attacks. This makes it crucial to adopt preventive measures that enforce security policies to restrict unauthorised access.

These security policies are implemented through Network Access Control (NAC) systems, which are particularly important for securing Local Area Networks (LANs), where users and devices connect to share resources. NAC systems ensure that only authenticated and authorised users or devices can access the network. This approach involves three main tasks: authentication, authorization, and accounting (AAA). 

Authentication, which verifies the identity of users or devices, is a critical step in securing enterprise networks. The most common standard for network access authentication is IEEE 802.1X, also known as Port-Based Network Access Control (PNAC). This standard is widely implemented in modern networks, including WPA2-Enterprise Wi-Fi (IEEE 802.11i) and Ethernet using MACsec (IEEE 802.1AE), providing secure access to both wired and wireless environments.

Although this paper focuses on IEEE 802.1X, several other technologies support authentication in corporate networks, particularly within Zero Trust Architecture (ZTA), which emphasises continuous verification and minimal trust assumptions. RADIUS (Remote Authentication Dial-In User Service) is widely used for centralised authentication, authorization, and accounting, ensuring each access request is validated independently. TACACS+, developed by Cisco, extends RADIUS by providing more detailed control over authorization and command execution, making it ideal for environments needing granular access control. Lastly, Kerberos uses tickets for secure, mutual authentication between clients and servers, making it a key element in scalable, Zero Trust-based enterprise networks.

\subsection{Extensible Authentication Protocol (EAP)}
In IEEE 802.1X, the authentication process relies on the Extensible Authentication Protocol (EAP) \cite{vollbrecht_extensible_2004}, an adaptable framework that supports multiple authentication methods. The most widely-used methods are listed below in order of increasing complexity:

\begin{enumerate}
	\item EAP-GTC: A simple method where credentials are sent in plaintext from a token card.
	\item EAP-MD5: This method provides minimal security via the MD5 hash function, and is a mandatory inclusion in EAP-compliant devices.
	\item EAP-LEAP: A Cisco proprietary method supporting dynamic WEP keys and key rotation, but with plaintext packet transmission.
	\item EAP-CHAP: Uses challenge-response mechanisms where the client proves knowledge of a shared secret.
	\item PEAP/EAP-TTLS: These protocols establish a secure TLS tunnel for protecting the client’s credentials, using inner EAP methods for authentication.
	\item EAP-PAP: Credentials are sent in plaintext but are usually encrypted within a EAP-TTLS tunnel.
	\item EAP-MS-CHAPv2: An improved challenge-response method developed by Microsoft, offering password encryption.
	\item EAP-TLS: Provides the strongest security with mutual authentication through TLS certificates for both client and server.
\end{enumerate}

Although EAP-TLS is recognized as the most secure authentication method, its adoption is hindered by the requirement for a Public Key Infrastructure (PKI) to issue client certificates, leading to high implementation costs. This challenge is further compounded in BYOD environments, where users must manually install certificates on their devices to gain network access. These factors contribute to its limited popularity despite its strong security. In contrast, methods like EAP-MD5, LEAP, MS-CHAPv2, and PAP are more popular as they rely on password-based authentication. While these methods are widely used, they are susceptible to password-related attacks, such as phishing and password stealing.

To enhance security of some EAP methods, PEAP and EAP-TTLS were developed as EAP methods that authenticate the server to the client and establish a secure TLS tunnel. Within this tunnel, other EAP methods, such as PAP or MS-CHAPv2, are used to authenticate the user. These protocols serve as outer EAP methods, while the user’s credentials are verified through inner EAP methods.

EAP-FIDO is designed as an inner EAP method that operates within PEAP or EAP-TTLS. The outer EAP methods follow a two-phase process:
\begin{itemize}
	\item Phase 1: Establish a secure, end-to-end TLS-encrypted tunnel between the EAP-Peer and the EAP-Server, during which the server is authenticated to the client.
	\item Phase 2: Authenticate the user using an inner EAP method—in this case, EAP-FIDO.
\end{itemize}

\section{EAP-FIDO protocol}
\label{sec:eap-fido}

This section describes the EAP-FIDO protocol: the architecture within the Extensible Authentication Protocol (EAP) and the message flow between the Supplicant and the Authentication Server (AS).

\subsection{The design approach}
\label{subsec:eap-fido:design}
The design of the EAP-FIDO protocol leverages FIDO2 as its core authentication mechanism, and takes advantage of the principles of the EAP framework modularity and interoperability. This makes EAP-FIDO interoperable, requiring only the configuration of a compatible EAP server and supplicants, without the need for changes to the existing network infrastructure. This ensures broad compatibility across diverse network environments, allowing the protocol to be deployed efficiently without modifying the underlying network deployment.

FIDO2’s challenge-response model introduces a modern secure, passwordless authentication. In EAP-FIDO, the client is prompted with a cryptographic challenge from the server, which is signed by the client’s private key and verified using the server’s knowledge of the corresponding public key.

A key design consideration is maintaining the integrity of this process while ensuring mutual authentication. To achieve integrity, the protocol tightly integrates FIDO2’s attestation and assertion models, guaranteeing that both client and server can verify the authenticity of the involved cryptographic materials. To guarantee the mutual authentication, the EAP-FIDO messages are tunnelled in PEAP/EAP-TTLS frames. Finally, security against man-in-the-middle attacks is reinforced by binding the challenge-response exchanges to the session context, ensuring they cannot be reused or intercepted.

Special emphasis is placed on performance optimization, minimising latency in the FIDO2 key exchange while adhering to the lightweight nature of EAP, making the protocol efficient and usable in a wide range of network environments.

\subsection{Protocol architecture}
\label{subsec:eap-fido:architecture}
The EAP-FIDO architecture follows a standard EAP environment structure, comprising three core components: (1) user equipment, (2) the access point, and (3) the authentication server.

In this setup, the user verifies their identity using a FIDO2 authenticator, which may be a FIDO2 security key or platform-based authenticator, within the user equipment. Authentication begins when the user equipment connects to the Access Point (AP), serving as an intermediary. At this point, the server and user equipment engage in a FIDO2 challenge-response authentication. Upon successful authentication, the AP authorises network access for the user equipment.

Therefore, the three main components are:

\begin{enumerate}
	\item \textbf{User Equipment (EAP Peer)}. The user device connects to a Port-Based Network Access Control (PBNAC) protected LAN, known in IEEE 802.1X as the Supplicant. In EAP-FIDO, the User Equipment will also work as a FIDO2 Client.
	\item \textbf{Ethernet Switch / Wireless Access Point (EAP Authenticator}). The entity that provides access to the LAN through a logical or physical port, protected by PBNAC. It may implement a co-located Authentication Server. However, usually it is in charge of forwarding EAP messages to the remote Authentication Server. In this design, we will consider the Authentication Server to be remote.
	\item \textbf{Authentication Server (EAP Server)}. The entity in charge of implementing the server-side logic of EAP authentication protocols. In EAP-FIDO, the Authentication Server will also work as a FIDO2 Server.
\end{enumerate}

\begin{figure*}[!ht]
	\centering
	\includegraphics[width=1.0\linewidth]{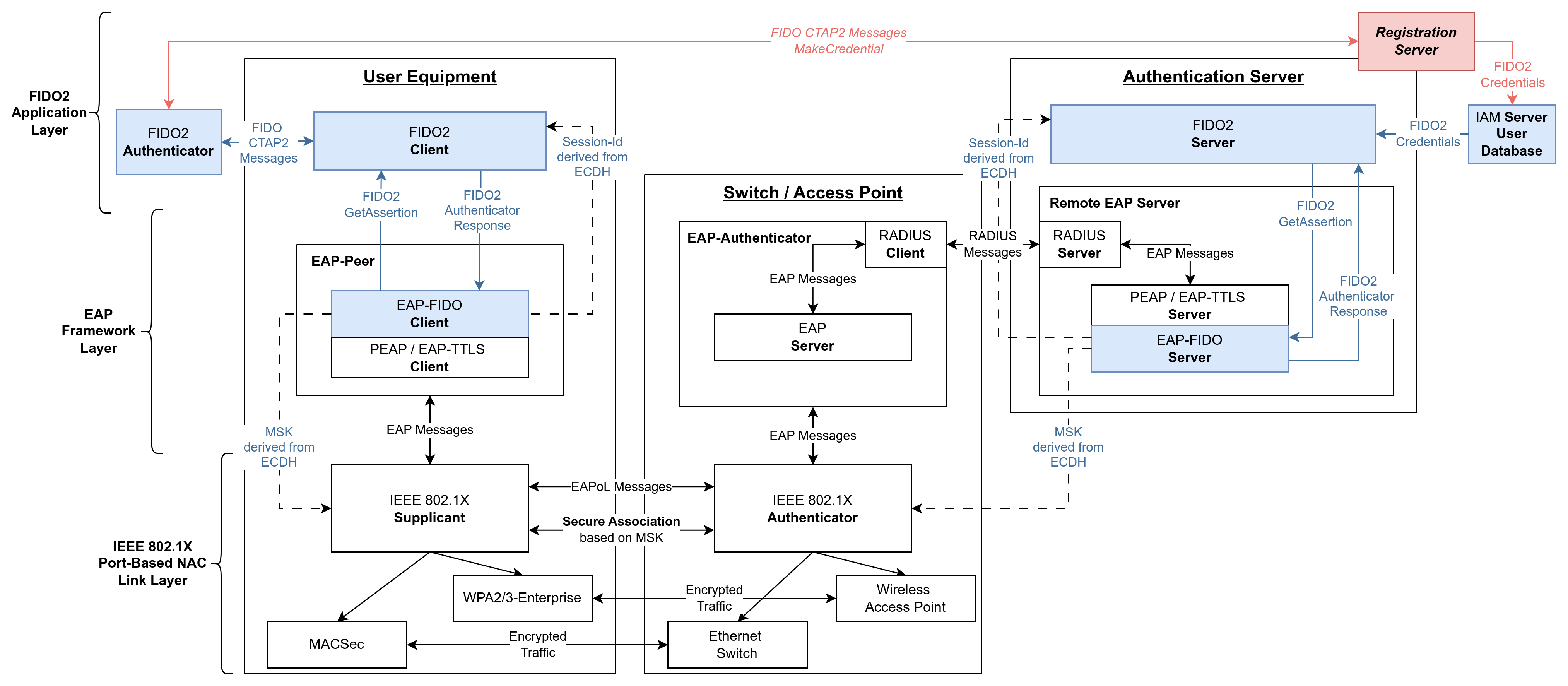}
	\caption{Architecture of the FIDO Extensible Authentication Protocol Method (EAP-FIDO).}
	\label{fig:eap-fido-architecture}
\end{figure*}

Figure~\ref{fig:eap-fido-architecture} illustrates the detailed system architecture of the three main elements and the flow of information between them. The figure also includes the registration server (shown in red), which, while part of the overall system, is not addressed by the EAP-FIDO design. The new or modified components specific to EAP-FIDO are highlighted in blue.

To ensure mutual authentication, EAP-FIDO messages are encapsulated within PEAP/EAP-TTLS TLS tunnels. This approach enables the EAP client to authenticate the EAP server before initiating the FIDO2 authentication, establishing a TLS-encrypted tunnel that guarantees the integrity and confidentiality of the exchanged messages. This secure tunnel protects the communication from potential eavesdropping or tampering, providing a robust layer of security before FIDO2 credentials are exchanged.

Figure~\ref{fig:eap-fido-architecture} presents the complete architecture in using a multiplexed view of the protocol. A detailed description of this figure follows below.

The left side of the figure shows the User Equipment, also referred to as the client. In EAP-FIDO, the client interacts with the FIDO2 authenticator, which the user connects during authentication. The FIDO2 client module handles this communication, sending low-level CTAP2 requests and processing the corresponding responses. It works with the EAP-Peer to encapsulate FIDO2 requests and responses into custom EAP-FIDO messages, which are then tunnelled within PEAP or EAP-TTLS frames. These encapsulated messages are finally transmitted by the 802.1X Supplicant module via Wi-Fi or Ethernet, using EAPoL frames.

At the centre of the figure, is the switch or access Point (AP), which receives the EAPoL frames and decapsulates them using the EAP-Authenticator module. This module is responsible for forwarding the EAP messages to the EAP server via RADIUS network-level communication. Importantly, EAP-FIDO requires no modifications to the EAP-Authenticator, ensuring that the protocol remains fully interoperable and easy to deploy in existing network environments that already support EAP.

Finally, on the right side of the figure is the Authentication Server. The RADIUS frames are decapsulated into EAP-FIDO messages, which are processed by the EAP-FIDO module. This module interacts with the FIDO2 server, responsible for verifying FIDO2 assertions using the credentials retrieved from the user database in the Identity and Access Management (IAM) server.

The FIDO2 challenge-response authentication is bound to the session context through the use of the Session-Id, as explained in section~\ref{subsec:eap-fido:operation} . Afterward, the 802.1X modules utilise the resulting Master Session Key (MSK) to derive cryptographic keys, which are used to secure network traffic following successful authentication.

EAP-FIDO is an Extensible Authentication Protocol (EAP) method. EAP was designed as a framework to implement new protocols to be used for Port-Based Network Authentication (PBNAC). 

EAP works on top of IEEE 802.1X, a standard that defines how to provide authentication, authorisation and key agreement mechanisms to support secure communication between devices of IEEE 802 Local Area Networks (LANs). For engineering coherence, EAP-FIDO design uses terminology of both EAP and IEEE 802.1X standards \cite{noauthor_ieee_2020} \cite{vollbrecht_extensible_2004}.

Here, we include the related standards of the multiplexed architecture of IEEE 802.1X systems for PBNAC. These are listed in a top-down abstraction layered approach:

\begin{itemize}
	\item FIDO2 Application Layer: FIDO CTAP2 \cite{armstrong_client_2022}
	\item EAP Framework Layer: EAP \cite{vollbrecht_extensible_2004}, RADIUS for EAP \cite{calhoun_radius_2003}, EAP-TTLS \cite{funk_extensible_2008} and PEAP \cite{palekar_protected_2004}.
	\item Port-Based Network Access Control Lower Layer: IEEE 802.1X, IEEE 802.1AE and IEEE 802.11. \cite{noauthor_ieee_2020}
\end{itemize}

\subsection{EAP-FIDO operation}
\label{subsec:eap-fido:operation}

EAP-FIDO is designed in three main stages, described below. For managing the EAP-FIDO messages, the protocol implements a state machine with OpCodes identifying each EAP-FIDO stage.

\begin{itemize}
	\item \textbf{Initialisation}: When initialising the EAP-FIDO method, the Supplicant may specify an EAP-Identity, which should be interpreted as the username in FIDO2. To continue, the state is changed to FIDO-Start in the AS. If the protocol uses a FIDO2 extension like HMAC-Secret for key derivation, then, the state is changed to FIDO-Request instead.
	
	\item\textbf{ FIDO-Start (opCode 1)}: The AS starts a ECDH key exchange with the Supplicant to share a secret for the MSK key derivation. To continue, the state is changed to FIDO-Request in the AS.
	
	\item \textbf{FIDO-Request (opCode 2)}: The AS generates a challenge that is sent to the Supplicant. The client searches for FIDO2 credentials and generates a FIDO CTAP2 get assertion request to the selected authenticator. The client should perform user verification (via a PIN) and user presence. To finish, the authenticator response is sent to the AS for validation.
\end{itemize}

\begin{figure*}[th!]
	\centering
	\includegraphics[width=1\linewidth]{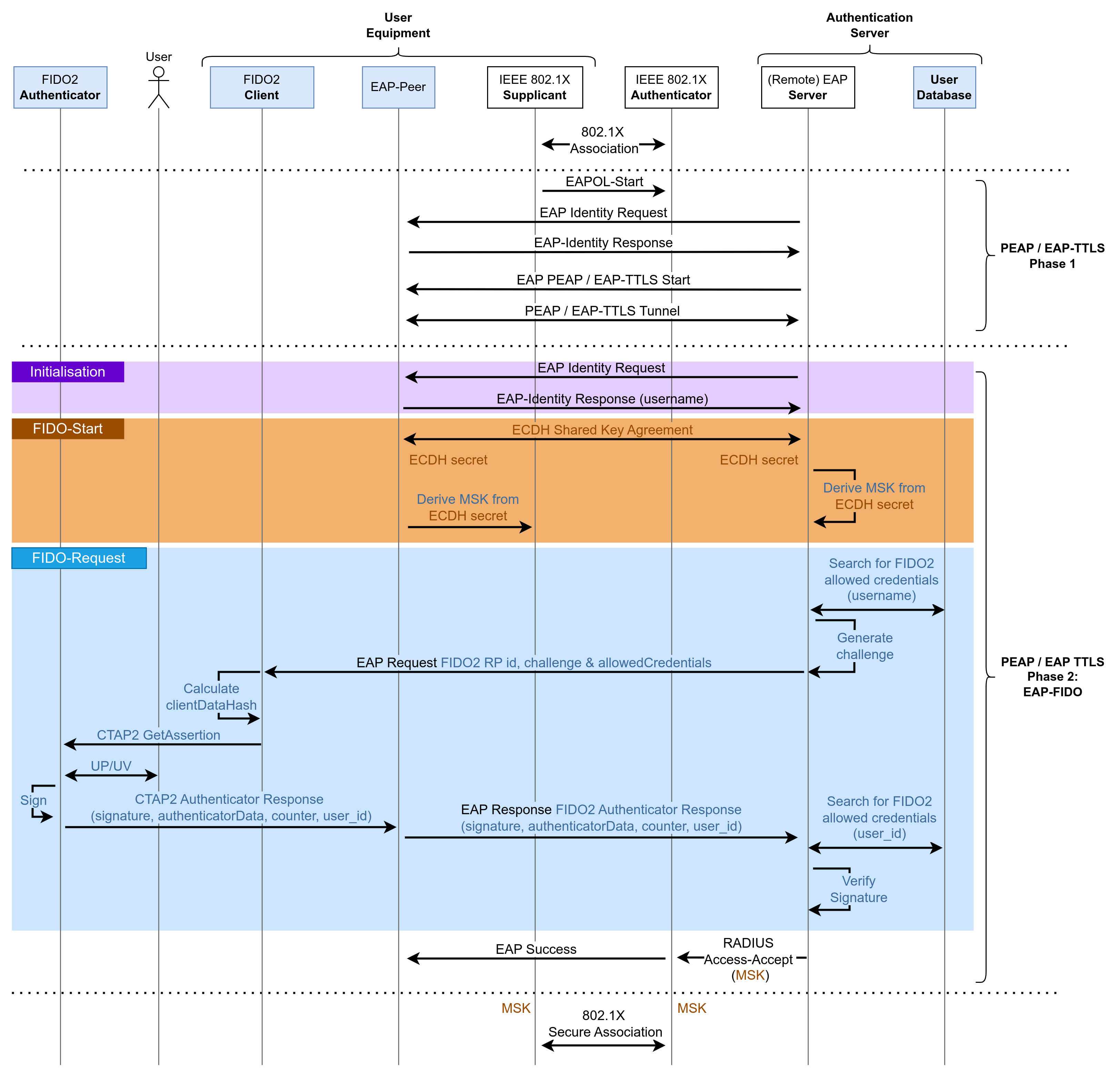}
	\caption{Message flow during EAP-FIDO Authentication encapsulated in PEAP / EAP TTLS.}
	\label{fig:eap-fido:flow}
\end{figure*}

Figure~\ref{fig:eap-fido:flow} represents the complete EAP-FIDO authentication message flow diagram, following the main EAP authentication process. All the stages of EAP-FIDO are tunnelled using PEAP / EAP-TTLS to ensure mutual authentication, as mentioned in section~\ref{subsec:eap-fido:design}. Notice that the message flow omits the details of the TLS handshakes, and assumes that EAP-Peer and EAP Server implement this encapsulation. Hereafter, a detailed protocol description is included. 

The protocol starts with the 802.1X Association. The Supplicant initiates the connection to the Authenticator, forwarding EAP packets to the EAP remote server. The EAP authentication is a PEAP / EAP TTLS two-phase protocol. In Phase 1, the tunnel is created via a TLS handshake. Phase 2 (the inner EAP method) is EAP-FIDO, where messages are tunnelled.

\subsubsection{EAP-FIDO: Initialisation}
The first message in an EAP method is the \textbf{EAP-Identity}. In EAP-FIDO, the EAP-Peer may include the username identifying the user in the EAP-Identity Response. When using discoverable credentials, this identity should be set to ``anonymous''.

\subsubsection{EAP-FIDO: FIDO-Start}
The Authentication Server (AS) initiates the \textbf{Elliptic Curve Diffie-Hellman (ECDH) key exchange} with the EAP-Peer, ending in an ECDH secret (\(K\)) that is used to derive the \(MSK\), as explained later.

\begin{figure*}[th!]
	\centering
	\includegraphics[width=1\linewidth]{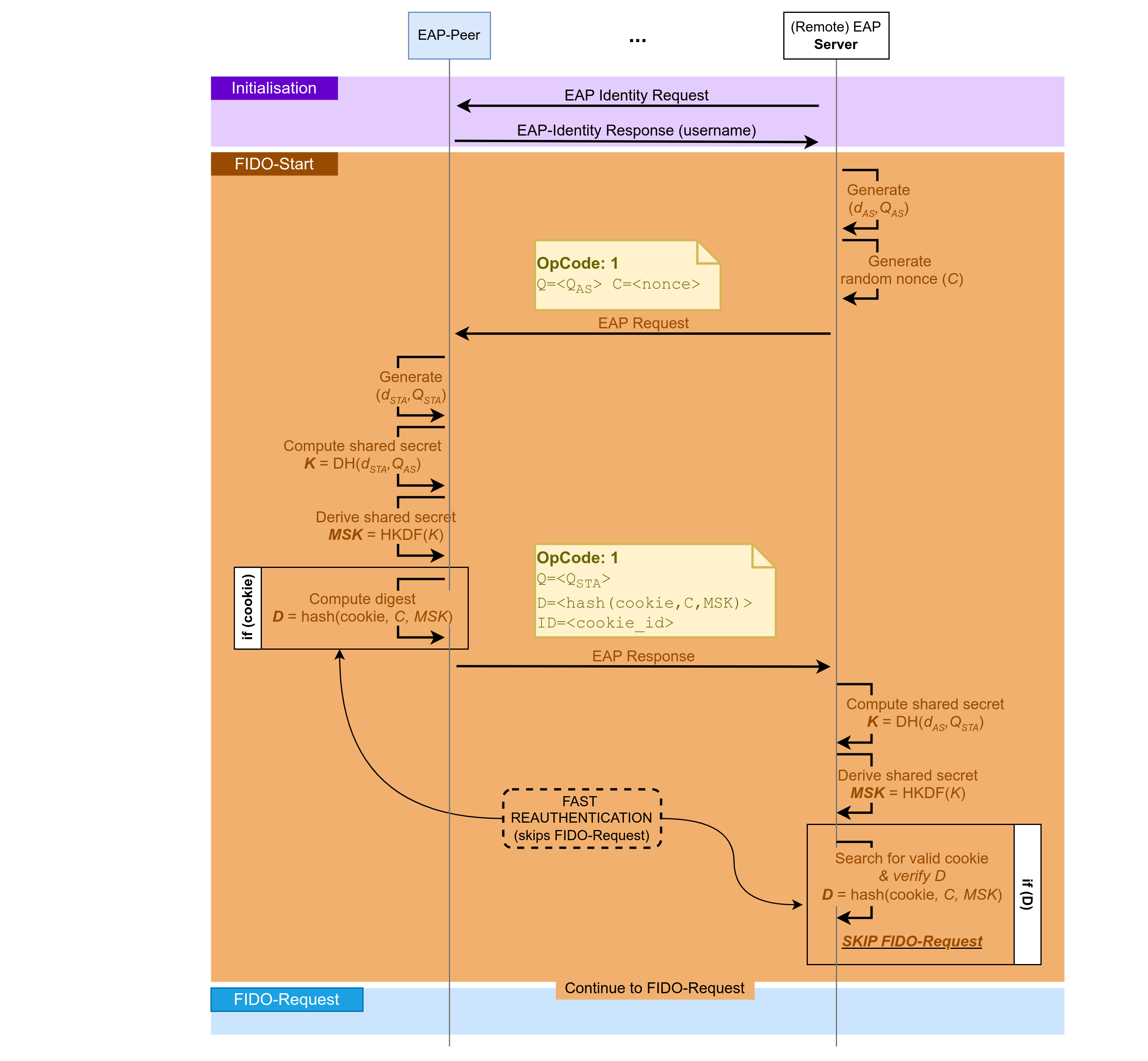}
	\caption{Message flow during EAP-FIDO Authentication encapsulated in PEAP / EAP TTLS.}
	\label{fig:eap-fido:start}
\end{figure*}

As seen in Figure~\ref{fig:eap-fido:start}, the AS creates a random private key (denoted \(d_{AS}\)) and generates its corresponding public key \(Q_{AS}\) using elliptic curve cryptography. This public key is then included as parameter \(Q\) in the following EAP Request. Additionally, the AS generates a nonce, which is included as the parameter \(C\).

When the EAP-Peer receives the EAP Request, it produces its own elliptic curve key pair (\(d_{STA}\), \(Q_{STA}\)) and computes a shared secret \(K\) based on its private key \(d_{STA}\) and the public key \(Q_{AS}\) from the AS. The EAP-Peer uses the HMAC-based Extract-and-Expand Key Derivation Function (HKDF) to derive 64 bytes for the Master Session Key (MSK) from this shared secret \(K\). To finalise the Diffie-Hellman key exchange, the EAP-Peer sends its public key \(Q_{STA}\) as parameter \(Q\) in the EAP Response.

For providing fast re-authentication, if the EAP-Peer still holds a session token from an earlier session, it may provide a proof of possession for this token within the EAP Response. The token includes a pair consisting of the Cookie Id and Session Cookie, which are stored locally by the EAP-Peer. To prove possession, the Supplicant calculates a digest \(D\) by combining the Session Cookie, the nonce \(C\) provided by the AS, and the \(MSK\). This proof of token possession is represented by the pair (\(\text{Cookie Id}, D\)). The digest \(D\) assures the AS of the token’s freshness, while also confirming the validity of the Session Key without revealing it to any potentially unauthorised AS.

Finally, once the AS receives the EAP Response, it extracts \(Q_{STA}\) and uses its private key \(d_{AS}\) to compute the shared secret \(K\), which it then uses to derive the \(MSK\) using HKDF. If token proof is provided, the AS retrieves the Cookie Id and checks it against its records of active sessions. If the session is valid and tied to an identity, the AS calculates a digest in the same way as the Supplicant and compares it to the received value \(D\). If the digests match, the AS bypasses the next step of EAP-FIDO authentication.

\subsubsection{EAP-FIDO: FIDO-Request}

\begin{figure*}[th!]
	\centering
	\includegraphics[width=1\linewidth]{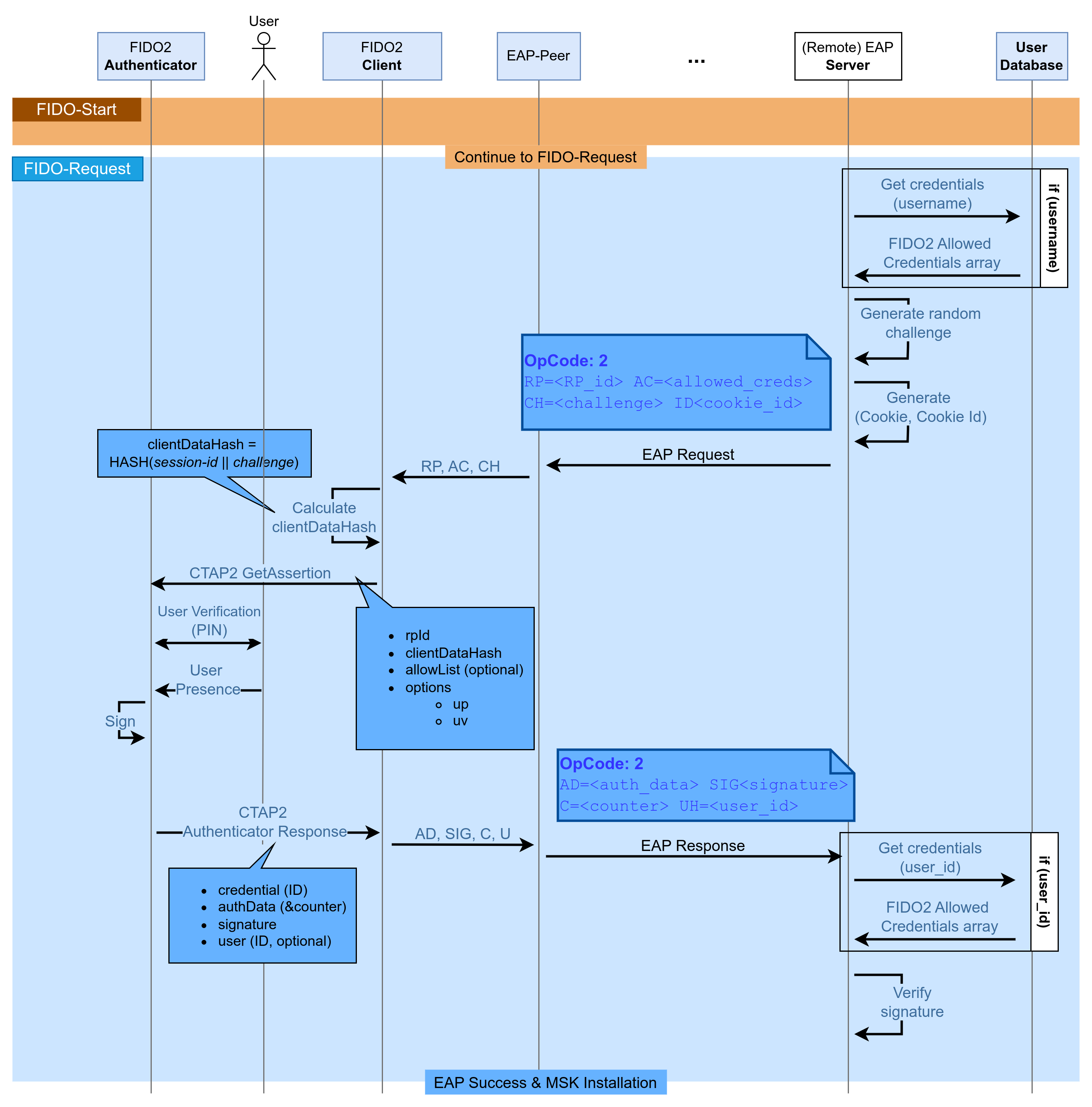}
	\caption{Message flow of EAP-FIDO. FIDO-Request stage.}
	\label{fig:eap-fido:request}
\end{figure*}

Figure~\ref{fig:eap-fido:request} represents the next protocol stage. The AS sends a challenge to the user FIDO2 authenticator, which issues a FIDO2 assertion to complete the authentication process. For configuring the \textbf{FIDO2 request} in the user equipment, the AS sends an EAP Request with the challenge, the Relying Party (RP) identifier, the list of allowed credentials and the cookie identifier. All these parameters are configured as follows.

The \textbf{challenge is generated randomly at the AS}, and the RP id is a constant value that represents the domain name of the AS. The RP id must be the same configured during the FIDO2 credential registration (attestation). If the username was included in the FIDO-Start messages, the \textbf{AS searches for the valid registered credentials} for the user. The list of allowed credentials is an array of strings that correspond with the FIDO2 credential identifiers. 

Additionally, the AS generates a Session Cookie that may later be associated with a User Identity if the user successfully completes the authentication process. This Session Cookie is a digest of the Master Session Key (MSK), which is already known by both the AS and the EAP-Peer. As a result, there is no need for the AS to transmit the actual Session Cookie to the Supplicant. Instead, it provides the Supplicant with an identifier (\textit{Cookie Id}), sent as the Data parameter ID. Using this, the AS could verify the EAP-Peer’s proof of possession of the session token without exposing sensitive information.

When the EAP-Peer receives the EAP Request, it extracts all the parameters and \textbf{calculates the clientDataHash}. This hash is a digest of the EAP Session-Id and the challenge sent by the AS, which will provide contextual binding for the next step: the FIDO CTAP GetAssertion call.

The \textbf{FIDO2 Client issues a CTAP2 GetAssertion request} to the authenticator. This call is sent to the user authenticator, including the Relying Party id, the list of allowed credentials, the clientDataHash and the user verification and user presence options activated. This process requires user verification. In physical security keys, this is usually a PIN introduced by the user in the FIDO-Client. In smartphones, the FIDO2 authenticator may be protected using biometric authentication. After the user verification and presence, the authenticator issues a\textbf{ CTAP2 Authenticator Response}.

The Authenticator Response includes a signature, the authenticator data (RP id hash and flags), the signature counter and a user handle (user identifier). All these parameters are included in the EAP Response sent to the AS. The authenticator data and the signature are encoded in base 64.

The AS receives the authenticator response from the EAP-Peer and starts the \textbf{FIDO2 signature verification} process. If an empty list of allowed credentials was sent in the request, the FIDO2 authentication is using discoverable credentials. That is, that the user handle (user identifier) received from the FIDO2 authenticator is used to search for the corresponding registered credentials. At this point, the signature counter received from the FIDO2 authenticator should be higher than the last registered one.

To verify the FIDO2 signature, the clientDataHash is calculated from the challenge and the EAP Session-Id, with the same operation as the FIDO2 client. With the clientDataHash and the public key of the FIDO2 credential, the received signature is verified with the corresponding public-key algorithm used during the credential registration.

After successful EAP-FIDO authentication, the EAP-Server notifies the EAP-Peer via the \textbf{EAP-Success} message. This will trigger the \textbf{MSK derivation from ECDH secret}. Here, both the EAP-Peer and EAP-Server derive the MSK from the shared ECDH secret, installing the MSK in the IEEE 802.1X Supplicant and Authenticator. Finally, the \textbf{802.1X Secure Association} is established. The Supplicant and the Authenticator derive encryption keys for the traffic using the MSK.

\subsubsection{EAP-FIDO packet structure}
In EAP-FIDO, the packet structure is defined with the following fields:

\noindent\\\textit{Standard EAP fields}
\begin{enumerate}
	\item \textbf{Code}: Request (\texttt{0x1}), Response (\texttt{0x2})
	\item \textbf{Identifier}: One-byte field to match response with requests.
	\item \textbf{Length}: Two-byte field with the total EAP packet length.
	\item \textbf{Type}: \texttt{20} (selected from the available values: IANA).
\end{enumerate}

\noindent\\\textit{EAP-FIDO non-standard fields}
\begin{itemize}
	\item \textbf{OpCode}: One-byte field with each specific EAP-FIDO stage:
	\begin{itemize}
		\item \texttt{0x1}: FIDO-Start
		\item \texttt{0x2}: FIDO-Request
	\end{itemize}
	\item \textbf{Data}: Contains every other content to achieve authentication.
	\begin{itemize}
		\item \textbf{FIDO-Start request}: \texttt{Q=<$Q_{AS}$> C=<nonce>}
		\item \textbf{FIDO-Start response}: \texttt{Q=<$Q_{STA}$> D=<hash(cookie,C,msk)> ID=<cookie\_id>}
		\item \textbf{FIDO-Request request}: \texttt{RP=<RP\_id> AC=<allowed\_creds> CD=<client\_data> ID=<cookie\_id>}
		\item \textbf{FIDO-Request response}: \texttt{AD=<auth\_data> SIG=<signature> C=<counter> UH=<user\_handle>}
	\end{itemize}
\end{itemize}

The EAP-FIDO Data packet field includes key-value pairs with different sizes. Each key-value pair is joined by the equal sign ("\texttt{=}") and separated from the other pairs by a single space. The client data, authenticator data and signature buffers are encoded in base 64 for transmission. The allowed credentials list is composed of each credential id (encoded in base 64) separated from the next value with a comma ("\texttt{,}") without spaces.

\subsubsection{Key agreement for IEEE 802.1X Secure Association}
EAP-FIDO is designed as an EAP method providing Master Session Key (MSK) key derivation from FIDO2 authentication \cite{simon_extensible_2008}. The MSK is later used for the IEEE 802.1X key agreement process during the Secure Association, with the objective of encrypting traffic between the Supplicant and the Authenticator. For instance, for the Ethernet MACSec key distribution or for the 4-way handshake of WPA2/3-Enterprise.

In EAP-FIDO, a Elliptic Curve Diffie-Hellman (ECDH) shared secret is used to derive the MSK. Here, we assume that both sides agree on an EC, but a negotiation would only imply further protocol rounds. The MSK is installed in the IEEE 802.1X Supplicant and Authenticator after the successful FIDO2 authentication.

Although in our proposal we have used ECDH for key agreement and derivation, there are other possibilities thanks to the FIDO2 protocol. It is possible to derive a MSK by taking advantage of FIDO2 extensions. Instead of an ECDH at the beginning of the message flow, EAP-FIDO can use the HMAC-Secret FIDO2 extension.

With this alternative, during registration, the HMAC-Secret shared secret is configured into the FIDO2 authenticator. Then, during authentication, the HMAC-Secret is challenged by the server, and the HMAC-Secret extension result is kept in the Supplicant for MSK derivation. In the server, the same HMAC operation is performed to obtain the same MSK.

Notice that, by using the FIDO2 extension, the FIDO-Start flow is not required, as the result of the extension would be returned during the authentication message flow.

Finally, to improve user experience, we propose the use of a fast re-authentication cookie. This approach, followed by other works like Mortágua et al. \cite{mortagua_enhancing_2024}, uses an optional fast re-authentication cookie that allows maintaining the EAP session alive. This removes the need to re-authenticate frequently, improving the user experience.

\subsubsection{Contextual binding with EAP-FIDO Session-Id}
In the WebAuthn specification \cite{bradley_web_2021}, both the Relying Party (the server) and the client include contextual bindings in the client data to be signed by the authenticator. This ensures that the signed data includes information about the authentication context, avoiding Man in the Middle (MitM) and replay attacks. EAP-FIDO may also implement contextual binding during FIDO2 authentication using the EAP Session-Id.

"The EAP Session-Id uniquely identifies an EAP authentication exchange between an EAP peer and a server" \cite{simon_extensible_2008}. The Session-Id may be defined by the specific EAP method. In EAP-FIDO, a Session-Id uniquely identifies the authentication exchange and is bound to the FIDO2 authentication process.

In this paper, we propose a specific Session-Id derivation from the ECDH key exchange during the FIDO-Start operation. Our definition mimics the proposal of the PEAP method \cite{dekok_extensible_2020}. In EAP-FIDO, the Session-Id derivation would be defined as:

\[
\text{Session-Id} = \texttt{0x20} \, || \, \texttt{as.pubkey} \, || \, \texttt{sta.pubkey}
\]

The Session-Id is defined as the concatenation of the EAP-FIDO type code (20) and the Authentication Server (AS) and Station (STA, also known as Supplicant) ECDH public keys (\(Q_{AS}\) and \(Q_{STA}\), respectively).

The contextual binding is achieved by including the pre-computed Session-Id in the FIDO Client Data collection together with the challenge from the server. This collection is hashed, producing the clientDataHash, which is signed by the FIDO2 authenticator. Our proposal does not use the JSON-compatible serialisation algorithm used in WebAuthn L2 \cite{bradley_web_2021}, which is intended to provide further functionality in web scenarios. Instead, our approach is a binary append of the Session-Id to the challenge:

\[
\text{Client-Data} = \texttt{session-id} \, || \, \texttt{challenge}
\]

The clientDataHash is the SHA-256 of Client-Data. The calculation of this value required during the GetAssertion operation in the client, and during the validation of the authenticator response in the AS.

\subsubsection{FIDO CTAP2 configuration}
There are two types of FIDO2 credentials regarding its storage: discoverable (client-side) and non-discoverable (server-side) credentials \cite{bradley_web_2021}. Discoverable credentials are the modern FIDO2 credentials that simplify the identification of the user. However, non-discoverable credentials guarantee backwards compatibility with FIDO2 authenticators with limited storage capacity.

EAP-FIDO can support both discoverable and server-side credentials. Discoverable credentials do not require an identity previous to the FIDO2 authentication, so using them in EAP-FIDO allows anonymous identity configuration. The user is identified at the moment of the authenticator response verification by using the user handle. This attribute returned by the authenticator after the signature corresponds to the user identifier.

In contrast, the server-side credentials require the identity of the user to be specified in the EAP-FIDO Supplicant configuration. During FIDO2 authentication, the AS server must send a list of allowed credentials (a list of credential identifiers) to the Supplicant.

The configuration options of FIDO CTAP2 assertion (authentication) operation is described below. Notice that all these fields are standard in the FIDO CTAP2 specification, which permit flexibility to support different scenarios. For EAP-FIDO configuration, there are two options: discoverable credentials (with EAP-FIDO anonymous identity) and server-side credentials (by previously identifying the user).

\noindent\\\textit{FIDO CTAP2 Get Assertion options}
\begin{itemize}
	\item \textbf{RP ID}: A Relying Party identifier, usually a valid domain string, sent in by the EAP-FIDO server. It must be the same as the RP ID configured during attestation (registration) of the user credential.
	\item \textbf{Client data hash}: A hash of the serialised client data. It is built on the EAP session-id and the challenge (a random nonce generated by the server). As it is different in every assertion call, this hash avoids replay attacks.
	\item \textbf{Allow list (server-side credentials)}: An array of credential identifiers is included only when server-side credentials are used. When using discoverable credentials, this array is empty. This list of allowed credentials corresponds to the registered credentials of the user to be authenticated. Therefore, the user should be identified in the EAP-FIDO identity field before the assertion operation, as described before.
	\item \textbf{Options (\texttt{up}, \texttt{uv})}: User verification (\texttt{uv}) and User presence (\texttt{up}) options should be activated (configured to "\texttt{true}"). The user presence requests consent to the user, and the user verification requests to unlock the FIDO2 authenticator (for example, by a PIN). As described in FIDO CTAP2.1, it may be required to execute the PIN/UV authentication protocol.
\end{itemize}

\section{Prototype implementation}
\label{sec:prototype-implementation}

We have implemented EAP-FIDO as a prototype in \texttt{hostap} software codebase, a well-known project for wireless networking software managed by W1.f1 \cite{malinen_hostapd_nodate}. Related work has also based their prototypes with \texttt{hostap} codebase \cite{malinen_hostapd_nodate}. This section provides technical insight about the implementation.

\subsection{Architectural design of the implementation}
\begin{figure*}[th!]
	\centering
	\includegraphics[width=0.8\linewidth]{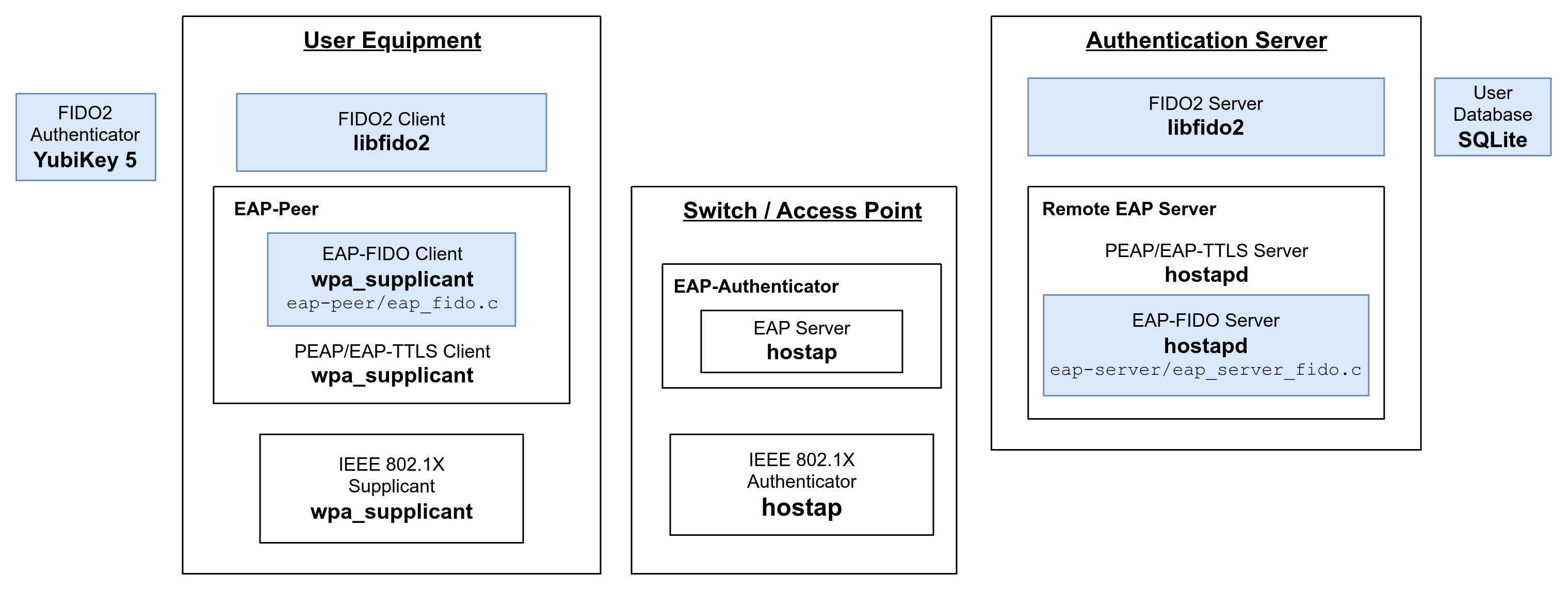}
	\caption{Architectural design of the EAP-FIDO prototype implementation.}
	\label{fig:implementation:architecture}
\end{figure*}

Figure~\ref{fig:implementation:architecture} shows the architectural design of our implementation, following the EAP-FIDO architecture shown in section 3.2. From left to right: (1) the user equipment, (2) the access point or switch, and (3) the authentication server at the right.

To implement both the FIDO2 client and server functionality, we use the libfido2 library. For the basic standard EAP Supplicant, Authenticator and Server functionality we rely on \texttt{hostap}. 

The \texttt{hostap} software codebase has a widespread use, it is compatible with several EAP methods and provides flexibility. As it implements wireless networking for both clients and servers, it provides a realistic network environment for analysing the effectiveness of EAP-FIDO.

Its server-side application is known as \texttt{hostapd} (Host Access Point Daemon), which allows a computer to act as an Access Point (AP). It supports network authentication by also implementing an Authentication Server (AS) for EAP. It supports several EAP methods and different WPA protocols. 

On the other hand, \texttt{wpa\_supplicant} is the application in charge of handling the network authentication on the client side, also known as Supplicant.

Figure~\ref{fig:implementation:architecture} highlights in blue the changes to the \texttt{hostap} source code, following the same changes proposed in the EAP-FIDO architecture. Notice that the EAP authenticator does not require any changes, while the EAP server and client are the main implementation targets. Therefore, the necessary implementations of the EAP interface (\texttt{eap\_i.h}) are in the client (\texttt{wpa\_supplicant}) and in the server (\texttt{hostapd}). Specifically, in this order:

\begin{itemize}
	\item EAP-FIDO Server (\texttt{eap-server/eap\_server\_fido.c}). It implements the server-side routines, the EAP header structure and the EAP session data structure.
	\item EAP-FIDO Client (\texttt{eap-peer/eap\_fido.c}). It implements the client-side routines,  and the same EAP header and session data structures as the client-side implementation
\end{itemize}.

The functions in the server implementation are called by the \texttt{process()} function, which processes EAP Responses taking into account the status of the state machine. Each of these routines manage the corresponding EAP Response, updates the state and builds the EAP Response.

However, the client implementation waits for EAP Requests and does not implement a state machine. Instead, it verifies the \textit{OpCode} to call the appropriate routine.

\subsection{Configuration in hostap}
There are several configuration three files in the hostap software: (1) the Access Point / EAP Server configuration in hostapd.conf; (2) the EAP server configuration in \texttt{eap\_users}; and  (3) the Supplicant configuration in \texttt{wpa\_supplicant.conf}. 

The example configuration of the server is included in the Listing~\ref{lst:implementation:hostapd}. If the wireless interface and driver is used, the server will make the computer act as a wireless AP. This example configuration works with a wired scenario, when hostapd is working as a wired IEEE 802.1X AP, EAP Authenticator and EAP Remote Server.

\begin{lstlisting}[caption=\texttt{hostapd.conf}, label=lst:implementation:hostapd]
# wired or wireless interface
interface=<eth0 or wlan0>
driver=<wired or nl80211>

# IEEE 802.1X configuration
ssid=EAP-FIDO-AP
ieee8021x=1

# EAP server configuration
eap_server=1
eap_user_file=../eap_users

# RADIUS server configuration
radius_server_clients=../clients
radius_server_auth_port=1812
radius_server_acct_port=1813

# Certs and keys
ca_cert=../certs/ca.pem
server_cert=../certs/server.pem
private_key=../certs/server.key
private_key_passwd=<secret key>
\end{lstlisting}

The referenced EAP server configuration is shown in Listing~\ref{lst:implementation:eap-users}. Here, the EAP identities are associated with the EAP method that should be used. All identities (including \textit{anonymous}) of Phase 1 are configured to use PEAP. In PEAP Phase 2, hostapd configuration does not allow the use of wildcards, so a copy of all identities (\textit{usernames}) should be configured here to use EAP-FIDO. The identity (\textit{username}) in EAP-FIDO is not mandatory, so "\texttt{anonymous}" / "\texttt{?}" is allowed here as anonymous identities.

\begin{lstlisting}[caption=\texttt{eap\_users}, label=lst:implementation:eap-users]
## PHASE 1
* PEAP

## PHASE 2
#Anonymous identity
"?"   	      FIDO    [2]
"anonymous"   FIDO    [2]

# Below list must be a copy of the users list in the DB.
"john-smith"  FIDO    [2]
"admin"       FIDO    [2]

\end{lstlisting}

Within the Supplicant configuration, shown in Listing~\ref{lst:implementation:wpasupplicant}, the client is configured to use the specified AP network, with PEAP using EAP-FIDO in Phase 2. The identity of PEAP is anonymous, and in "\texttt{identity}" a username to identify the user when using non-discoverable credentials. When using discoverable credentials, the identity should be "\texttt{anonymous}" or "\texttt{?}", which is equivalent to not specifying a specific identity (see section~\ref{subsec:eap-fido:operation}).

\begin{lstlisting}[caption=\texttt{wpa\_supplicant.conf}, label=lst:implementation:wpasupplicant]
# ...
network={
	ssid="EAP-FIDO-AP"
	key_mgmt=<IEEE8021X or WPA-EAP>
	eap=PEAP
	anonymous_identity="anonymous"
	identity="?"
	phase1="peaplabel=0"
	phase2="auth=FIDO"
}
\end{lstlisting}

\subsection{EAP-FIDO CTAP2 with \texttt{libfido2}}
In our implementation, we have used \texttt{libfido2} for interacting with the FIDO2 authenticator (i.e. a security key) via FIDO CTAP2. \texttt{libfido2} is maintained by Yubico and has widespread use nowadays \cite{martelletto_libfido2_nodate}. We have used the functionalities both in the client and in the server:

\begin{itemize}
	\item \texttt{fido\_get\_assert()}. When the Supplicant receives the challenge from the server, it configures and issues the assertion operation, which will return the FIDO2 signature and the FIDO2 authenticator data. We configured the operation with:
	\begin{itemize}
		\item 	\textit{Client data}: calculated using the challenge received from the server and the Session-Id of the EAP-FIDO session (see section~\ref{subsec:eap-fido:operation}).
		\item 	\textit{Relying party}: received from the server.
		\item 	\textit{User presence and user verification}: both are enabled.
		\item 	\textit{Timeout}: 30 seconds.
		\item 	\textit{PIN}: requested via the UI to the user to unlock the authenticator.
	\end{itemize}
	\item \texttt{fido\_assert\_verify()}. Once the AS receives the authenticator response from the Supplicant, it is verified with this function using the user credential public key. The operation is configured as follows:
	\begin{itemize}
		\item 	\textit{Client data}: calculated using the challenge previously sent from the server and the Session-Id of the EAP-FIDO session (see section~\ref{subsec:eap-fido:operation}).
		\item 	\textit{Relying party}: the RP identifier  configured in the AS settings, which corresponds to the same origin of the RP during the registration.
		\item 	\textit{User presence and user verification}: both are enabled.
		\item 	\textit{Signature}: the signature received from the Supplicant.
		\item 	\textit{Public key}: the credential public key of the user authenticator, retrieved from the database.
	\end{itemize}
\end{itemize}

The signature counter that the AS receives should be greater than the one stored in the database. This condition is checked before verifying the FIDO2 signature.

In this paper, we do not focus on the registration operations. However, to test our implementation, we have developed a small code snippet using \texttt{libfido2} attestation operations to create and register the FIDO2 credential in the FIDO2 authenticator (i.e. the security key). We have also tested registering through a demo WebAuthn web application, and configuring the origin as the EAP-FIDO RP id in the AS.

\subsection{Authentication Server database}
\begin{figure}[h!]
	\centering
	\includegraphics[width=0.8\linewidth]{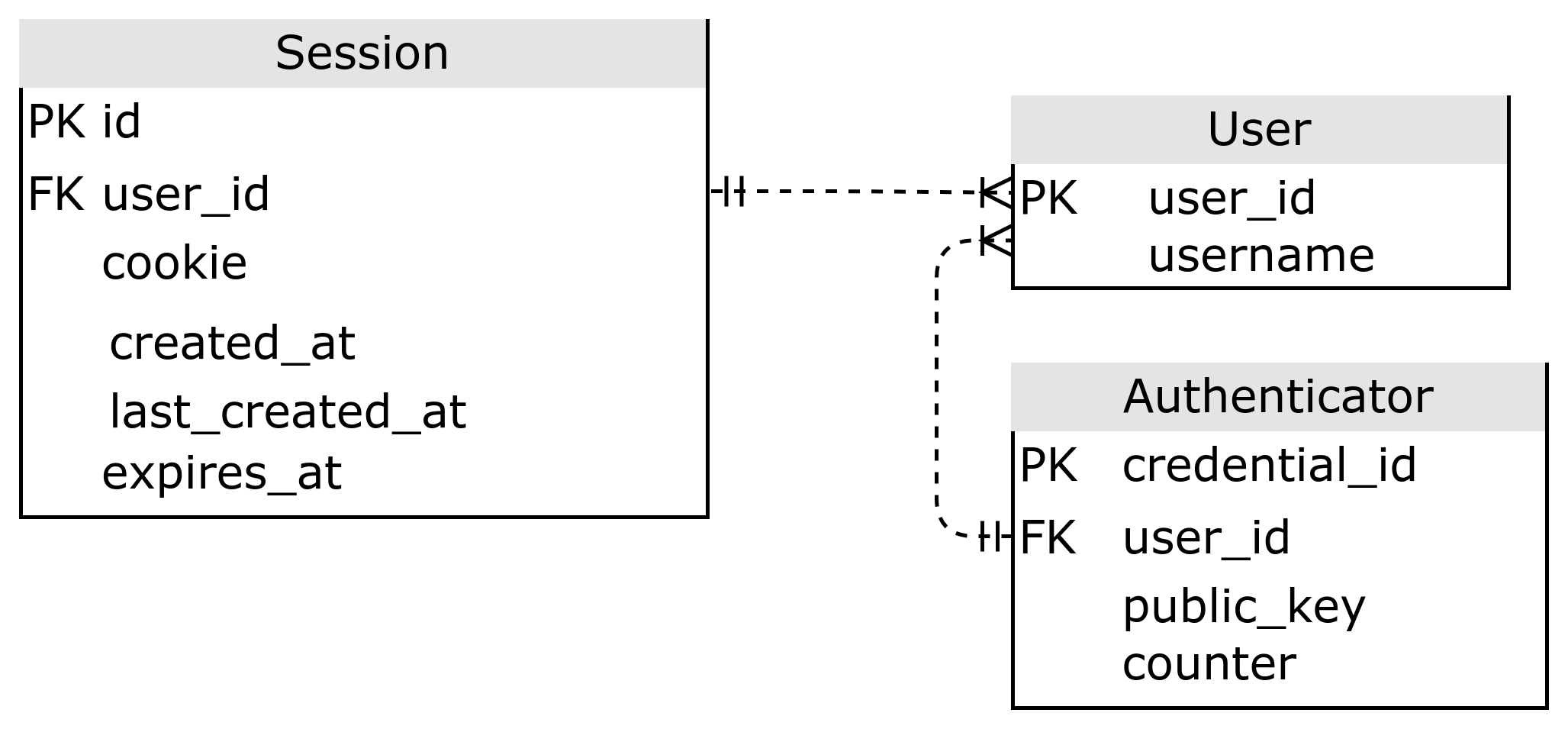}
	\caption{Database entity relational diagram of the implementation of EAP-FIDO in \texttt{hostapd}.}
	\label{fig:implementation:database-design}
\end{figure}

For this implementation, we created a relational database with 3 tables: "\texttt{session}", "\texttt{user}" and "\texttt{authenticator}". The registered users are identified by the \texttt{user\_id}, and they may have several associated FIDO2 credentials (here named as authenticator). Associated usernames should also be unique. Sessions are used to keep the cookie associated with the fast re-authentication mechanism. Figure~\ref{fig:implementation:database-design} depicts the entity relational diagram of the database. The FIDO2 challenge and other session data are stored in the EAP-FIDO session private data on memory.

\subsection{Cryptographic details}
During the FIDO-Start operation, an ECDH exchange was implemented. In this implementation, we have chosen the NIST P-384 curve \cite{chen_recommendations_2023}. As a hash algorithm, we selected SHA-256. Also, for simplicity, we have chosen to use EDDSA credentials during the FIDO2 registration, so we supported them during authentication. However, EAP-FIDO is not tight to a single type of FIDO2 credential.

\section{Performance analysis and results}
\label{sec:performance-analysis-and-results}

For validating EAP-FIDO, we have configured two different deployments of the prototype: a virtual wired network, and a Wi-Fi network. 

Then, the performance of EAP-FIDO has been measured based on time.

\subsection{Virtual wired network deployment}
During development, two virtual machines (VMs) were configured in a virtual network on VirtualBox 7.0. In this scenario, \texttt{hostapd} was working both as an Authentication Server (AS) and an EAP Authenticator, or Access Point (AP), hosted in one VM. The client side is executed with the wired driver of \texttt{wpa\_supplicant}. Both parts were connected using a Host-only adapter in VirtualBox. Finally, a Yubikey Security Key was used via USB through VirtualBox in the Supplicant VM.

\subsection{Wi-Fi network deployment}
\label{subsec:performance:wi-fi}
Once the virtual configuration was working, we deployed the solution in a real Wi-Fi environment. Apart from the AS and the Supplicant machines, this involves the usage of a Wi-Fi router (ASUS AC1200\_v2) and a wireless network card for the Supplicant (TP-link TL-WN722NV2).

\begin{figure*}[!th]
	\centering
	\includegraphics[width=0.6\linewidth]{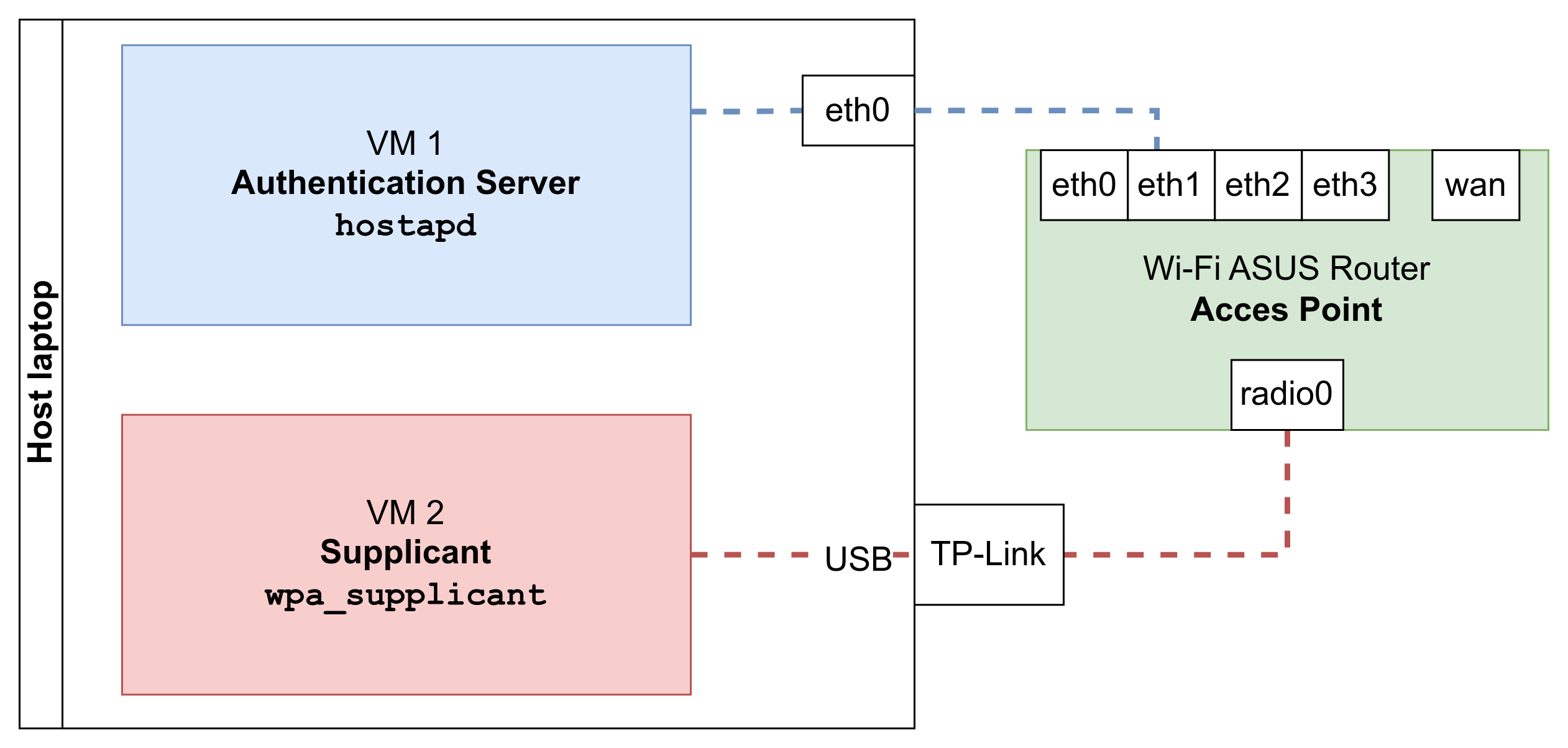}
	\caption{Wi-Fi deployment of EAP-FIDO.}
	\label{fig:performance:deployment-diagram-wi-fi}
\end{figure*}

Figure~\ref{fig:performance:deployment-diagram-wi-fi} shows the Wi-Fi deployment. It is important to mention that the usage of the same VMs for AS and for Supplicant than in the virtual network deployment has been done for consistency in the performance analysis. However, these can be understood as independent computers, one acting as a server and the other as a client. For the Supplicant VM to be able to access the Wi-Fi network, the TP-Link USB network card was plugged in through VirtualBox, like the Yubikey. Finally, the AS VM is connected to the router through the host ethernet interface.The configuration for this deployment involves some changes. The \texttt{hostapd} server now only acts only as an AS exposing the RADIUS ports to the ASUS AP. The Wi-Fi AP in ASUS is configured to work with WPA2-Enterprise using the external RADIUS server. Finally, the \texttt{wpa\_supplicant} is executed with the wireless drivers (option \texttt{-D nl80211}) and the corresponding interface. Also, its configuration now should use the “WPA-EAP” key management attribute.

\subsection{Performance analysis}
For testing the performance, we have implemented some timestamps in the EAP-FIDO \texttt{wpa\_supplicant} code, using the standard UNIX timestamp in microseconds (based on the \texttt{time.h} C library). We have used these timestamps to calculate the total elapsed time of the PEAP method, and the independent elapsed time of EAP-FIDO, following the methodology of Mortágua et al. \cite{mortagua_enhancing_2024}.

For this purpose, we established three timestamps: P1, recorded at the beginning of PEAP Identity (source file \texttt{eap.c}); P2 at the beginning of EAP-FIDO Identity (source file \texttt{eap\_fido.c}); and finally P3, after issuing EAP Success message in \texttt{eap.c}. After this, the three timestamps are written to a file in a single write operation. Until this moment, the timestamps are kept in memory to avoid additional I/O overhead in the measurement. Finally, the user interaction has been avoided, to specifically measure the EAP-FIDO performance. This was achieved by hardcoding the Yubikey PIN (user verification) and configuring FIDO to avoid user presence, so no touch is required form a human.

Table~\ref{tab:performance} shows the results of the performance analysis in milliseconds (N=20). We measured three cases: (1) authentication with server-side credentials; (2) authentication with discoverable credentials; and (3) fast re-authentication, implemented with the cookie. These measurements were taken in the Wi-Fi network deployment (see section~\ref{subsec:performance:wi-fi}) with the following characteristics:

\begin{itemize}
	\item \textbf{Host laptop}: MSI GL63 8RC, 16GB RAM, i7 8th Gen (12 cores), Manjaro Linux with VirtualBox 7.0.
	\item \textbf{VM1 AS}: 2GB RAM, 2 CPU cores, Ubuntu Server 22.04 LTS.
	\item \textbf{VM2 Sup.}: 6GB RAM, 4 CPU cores, Ubuntu Desktop 22.04 LTS.
	\item \textbf{FIDO security key}: Yubikey Security Key NFC, Firmware 5.4.3.
	\item \textbf{Wi-Fi Router}: ASUS AC1200\_v2, 64 MB RAM, OpenWRT 23.05.0.  
\end{itemize}

\begin{table*}[!th]
	\centering
	\begin{tabular}{l|rrrr|rrrr|}
		\cline{2-9}
		\multicolumn{1}{c|}{\textbf{}}                    & \multicolumn{4}{l|}{\textbf{EAP-FIDO (ms)}}                                                                                                     & \multicolumn{4}{l|}{\textbf{PEAP TOTAL (ms)}}                                                                                                   \\ \cline{2-9} 
		\multicolumn{1}{c|}{\textbf{}}                    & \multicolumn{1}{l|}{\textbf{avg}} & \multicolumn{1}{l|}{\textbf{stdev}} & \multicolumn{1}{l|}{\textbf{max}} & \multicolumn{1}{l|}{\textbf{min}} & \multicolumn{1}{l|}{\textbf{avg}} & \multicolumn{1}{l|}{\textbf{stdev}} & \multicolumn{1}{l|}{\textbf{max}} & \multicolumn{1}{l|}{\textbf{min}} \\ \hline
		\multicolumn{1}{|l|}{\textbf{Server-Side  Cred.}} & \multicolumn{1}{r|}{555.480}      & \multicolumn{1}{r|}{19.842}         & \multicolumn{1}{r|}{602.956}      & 525.758                           & \multicolumn{1}{r|}{791.019}      & \multicolumn{1}{r|}{37.649}         & \multicolumn{1}{r|}{858.426}      & 740.900                           \\ \hline
		\multicolumn{1}{|l|}{\textbf{Discoverable Cred.}} & \multicolumn{1}{r|}{521.297}      & \multicolumn{1}{r|}{26.006}         & \multicolumn{1}{r|}{574.934}      & 472.288                           & \multicolumn{1}{r|}{728.009}      & \multicolumn{1}{r|}{36.928}         & \multicolumn{1}{r|}{798.425}      & 663.281                           \\ \hline
		\multicolumn{1}{|l|}{\textbf{Re-authentication}}  & \multicolumn{1}{r|}{44.802}       & \multicolumn{1}{r|}{4.157}          & \multicolumn{1}{r|}{54.250}       & 38.135                            & \multicolumn{1}{r|}{255.786}      & \multicolumn{1}{r|}{22.755}         & \multicolumn{1}{r|}{323.387}      & 229.196                           \\ \hline
	\end{tabular}
	\caption{Performance results in miliseconds of the EAP-FIDO Wi-Fi deployment. }
	\label{tab:performance}
\end{table*}

Without user interaction, EAP-FIDO achieves an authentication time of 0.56 seconds ($\pm 0.02s$) using server-side credentials and 0.52 seconds ($\pm 0.03s$) with discoverable credentials (see Table~\ref{tab:performance}). The results show that EAP-FIDO performs similarly with both types of credentials, with discoverable credentials being 6.15\% faster.

When comparing EAP-FIDO to the overall PEAP process, we find that Phase 2 of PEAP accounts for roughly 70\% of the total authentication time. PEAP adds approximately 40\% overhead to the authentication process, as EAP-FIDO functions as an inner protocol within PEAP (see section~\ref{subsec:eap-fido:design}). This additional overhead is mainly due to the need for establishing the TLS tunnel in PEAP, which slows down the overall authentication compared to running EAP-FIDO as a standalone protocol.

In contrast, fast re-authentication requires only the initial round of the EAP-FIDO process, where the Supplicant provides a valid cookie to the Authentication Server (AS). This method is 91.93\% faster than server-side credential authentication, as it bypasses repeated user authentication within the same session, significantly reducing user interaction overhead.

FIDO2 authentication introduces a time cost  notable  when integrated with EAP but cannot be directly compared to other authentication methods. FIDO2 relies on external authenticator devices, which often interact with additional hardware, resulting in authentication delays. However, our EAP-FIDO prototype successfully completes communication, authentication, and verification in a Wi-Fi environment within an average of 0.6 seconds, regardless of the specific FIDO2 credential used.

This study did not focus on user interaction, as it largely depends on the design of the User Interface (UI). However, EAP-FIDO introduces a fast re-authentication token to minimise UI-related delay and maintain session continuity. Our protocol proposal suggests that EAP-FIDO configuration should be optimised to balance the frequency of user interactions with the duration of the re-authentication token's validity.

In brief, while FIDO2 integration into EAP introduces some overhead, the EAP-FIDO prototype demonstrates that secure and efficient authentication can be achieved in under a second in realistic Wi-Fi environments. Also, the implementation of fast re-authentication tokens offers a practical solution to reduce user interface delays, ensuring a seamless user experience without compromising security. Properly configuring this authentication method, makes EAP-FIDO a viable and scalable solution for secure network access.

\section{Security analysis}
\label{sec:security-analysis}

EAP-FIDO enables the use of FIDO2 authentication in network scenarios. Here, we analyse the security of EAP-FIDO, which was designed as an inner (tunnelled) EAP method in PEAP or EAP-TTLS. Notice that EAP-FIDO must be used as a tunnelled method to provide TLS server authentication to the supplicant. 

Therefore, in this section we analyse the security of the tunnelled EAP-FIDO. Our threat model takes into account attacks to the user credentials, Man in the Middle (MitM), rogue Authentication Server (AS), Evil Twin, replay and network sniffing attacks. These attacks to wireless networks can be used nowadays in WPA2/3-Enterprise networks \cite{palama_attacks_2023}. In this threat model, we do not consider host-based malware attacks to the Supplicant.

\subsection{Confidentiality and integrity}
A tunnelled EAP method like FIDO-EAP is protected by a TLS connection with PEAP or EAP-TTLS. The outer EAP method establishes a TLS connection after authenticating the AS to the supplicant, and encrypting the following EAP messages. This ensures not only confidentiality, but also integrity for the EAP messages.

\subsection{Evil Twin and the security of FIDO2 user credentials}
FIDO2 is a challenge-response authentication protocol using public-key cryptography. The FIDO2 user credentials reside in the user device or in the user hardware security key, protected in a secure storage. In this threat model, we do not consider malware attacks in the Supplicant, and we assume the UF-t weak security for FIDO CTAP2, as defined by Barbosa et al. \cite{malkin_provable_2021}.

FIDO2 credentials are never transmitted, as they are used to sign the corresponding challenge from the server. In the event of an Evil Twin (rogue Access Point) that issues a fake challenge and intercepts a valid signature does not compromise the credentials. Therefore, Evil Twin attacks to steal user credentials do not work with EAP-FIDO, unlike with other WPA2/3-Enterprise configurations based on password credentials \cite{palama_attacks_2023}.

On the other hand, the Authentication Server (AS) only stores the public keys of FIDO2 credentials, which are used to verify the signatures. An attack that compromises the AS could lead to privacy implications related to the username or other attributes that are stored. However, such a compromise does not leak FIDO2 credentials that allow the user impersonation.

\subsection{Rogue AS, MitM and replay attacks}
A rogue AS may impersonate the valid AS in EAP-FIDO, enabling attacks like Man in the Middle (MitM) and message replay. However, the design of EAP-FIDO mitigates these attacks.

A MitM attack to EAP-FIDO could hijack the ECDH initial exchange, by placing the attacker’s public key instead of the AS public key value. That would make the attacker be able to create a new IEEE 802.1X Secure Association between them and the Supplicant, performing a MitM.

To prevent this MitM attack, the EAP-FIDO AS and Supplicant derive a \textit{Session-Id} from the ECDH key exchange. The Supplicant includes the \textit{Session-Id} in the signed Client Data (see section~\ref{subsec:eap-fido:operation}). When the real AS verifies the signature from the FIDO2 authenticator, it will expect the calculated \textit{Session-Id} in the signature. If an attacker attempts to hijack the ECDH, the \textit{Session-Id} would be modified, which will cause the real AS failing to verify the FIDO2 signature. Therefore, it will abort the authentication.

On the other hand, an attacker may inspect the traffic \cite{palama_attacks_2023} to retrieve a valid signature from the FIDO-EAP packets to later perform a replay attack. That is, use the valid signature to perform authentication in a future authentication. EAP-FIDO mitigates replay attacks by the use of a random FIDO2 challenge (or nonce), different in each authentication to a signature invalid in a different FIDO2 authentication.

\section{Use cases}
\label{sec:use-cases}

This section includes some use cases for EAP-FIDO  network authentication, considering the current adoption and market perspectives of FIDO2 authentication.

\subsection{Corporate networks}
Enterprises and the different organisations may benefit from EAP-FIDO in the same way they do from the adoption of FIDO2 authentication (also known as passkeys authentication) in web applications. 

\begin{itemize}
	\item \textbf{Single FIDO2 credentials}: Employees using passkeys or security keys for accessing corporate web applications could directly use them to get access to corporate networks with EAP-FIDO. An organisation may register and associate FIDO2 credentials to an user via a web application, and then use them for network authentication by using EAP-FIDO.
	\item \textbf{Passwordless secure authentication}: Enterprises seeking a passwordless future that mitigates the risks associated with passwords will be able to also use FIDO2 in networks. This mitigates the risk of stolen credentials, and easies the path towards the complete removal of the password as an authentication method in corporate environments.
\end{itemize}

\subsection{Public hotspots}
Organisations with public hotspots for guests or clients can adapt EAP-FIDO to improve security, privacy and user experience.

\begin{itemize}
	\item \textbf{Easy registration and authentication}. Registering FIDO2 credentials like device-stored passkeys in smartphones involves few user clicks. Guests may be registered simply by accessing a web application by scanning a QR. They are not required to set up a new password or use other personal information. Later, they use the registered passkey to access the Wi-Fi network securely and in a fast way.
	\item \textbf{Improved security}: Using EAP-FIDO authentication makes access to networks to implement a more reliable control than captive portals deployed in open wireless networks, for example. Also, the risk associated with shared or stolen credentials is eliminated, mitigating user impersonation.
\end{itemize}

\section{Related work}
\label{sec:related-work}

This section includes the related work published up to date. EAP-FIDO integrates the FIDO2 authentication in networks, so the linked works are framed into these two areas. 

\subsection{FIDO2 authentication}
In FIDO2CAP, we have introduced FIDO2 in network authentication, using a WebAuthn-based web captive portal \cite{rivera-dourado_novel_2024}. The main difference with the EAP-FIDO protocol is that FIDO2CAP is not compatible with the standard 802.1X authentication, so it does not provide key derivation to encrypt the communication. Also, it relies on an enforcement device (firewall) at the network level to control the connectivity to network resources.

There are other works that have brought passwordless authentication to related scenarios, like the Internet of Things (IoT). Dixit et al. \cite{silhavy_fido2_2023} have implemented a method based on QR codes to use FIDO2 authentication in smart TVs, delegating it in a web application accessed through a smartphone. Authenticating IoT devices with FIDO is also supported by the work of Luo et al. \cite{luo_g2f_2021}, who use hardware FIDO tokens for verifying the IoT device identity in a gateway. Similarly, there are other works that even propose the use of FIDO authentication in hardware locks \cite{chakkaravarthy_sethuraman_loki_2022}, for users unlocking their rooms in Smart Hotels \cite{dammak_secure_2020}.

However, the main approach to integrate FIDO2 in networks has been the provisioning of temporal legacy credentials to users after the FIDO authentication flow. This is the strategy used in the work of Conners et al. \cite{conners_lets_2022}, where they provision user certificates after successful FIDO2 authentication. Similarly, Huseynov \cite{huseynov_passwordless_2022} has designed a middleware module that provides temporal passwords to a VPN client after FIDO2 authentication.

Provisioning temporal credentials has also been the approach of Chifor et al. \cite{chifor_flexible_2018}. They presented a solution that makes use of the FIDO UAF protocol, a protocol of the first generation of FIDO \cite{angelogianni_how_2024}. In their work, the protocol is used for authenticating guest users in an enterprise network by employing their personal Android smartphones. For this purpose, they have created a scheme where guests can register themselves to be authenticated in a different Wi-Fi network. Their approach, however, differs from EAP-FIDO. Their solution uses an already designed authentication mechanism existing in network access control, the EAP-PEAP method with MSCHAPv2, a password-based authentication method. Similarly to Huseynov \cite{huseynov_passwordless_2022}, the server provides a temporary pair of usernames and passwords, after the FIDO UAF authentication.

The approach of EAP-FIDO, described in this paper, is very different. Taking advantage of the EAP framework, EAP-FIDO directly verifies the user identity with FIDO2, without relying on other legacy credentials.

Recently, two proposals were published outside the scientific journals that implement FIDO credentials for accessing the network. The master thesis presented by Panizza \cite{panizza_fido2_2024}, where they propose a solution that embeds the FIDO authentication within the TLS handshake, resulting in a new TLS library which they use with EAP-TLS for network authentication. In any case, their approach may require further compatibility and security analysis.

On the other hand, Rieckers and Winter \cite{rieckers_eap-fido_nodate} have proposed a first draft of the EAP-FIDO protocol to the EAP Method Update (EMU) working group of the IETF. Although their specification is still under development, some of their design strategies are similar to the proposal in this paper. In their design, EAP-FIDO also works as an inner authentication protocol, supported by a TLS tunnel: first, a TLS handshake, and then the FIDO authentication.

However, there are key main differences in the design strategy of the EAP-FIDO protocol, further than the message format. Rieckers proposes the use of silent authentication, without user presence or verification. Although more usable, it may not comply with the standards and cause security risks. Also, their design of the protocol operation is different: they require additional protocol rounds for FIDO server-side credentials. Finally, there is not yet a complete definition of the key derivation in their proposal. In our paper, we describe a complete key derivation and contextual binding for the protocol stack in 802.1X authentication.

Several studies have examined the security and performance of FIDO2. Theoretically, Barbosa et al.~\cite{malkin_provable_2021} introduced a security framework to assess the robustness of these standards. In practice, tools like the one presented in~\cite{vasileios_grammatopoulos_web_2021} have been used in WebAuthn scenarios to analyze commercial and open-source implementations, uncovering critical misconfigurations~\cite{grammatopoulos_blind_2022}. More closely aligned with the objectives of our proposal, related works such as ~\cite{mortagua_enhancing_2024} and ~\cite{marques_eap-sh_2020} have conducted comprehensive evaluations of both performance and security.

\subsection{Differences with other EAP methods}
There are other recent proposals of EAP methods in the scientific literature, like the work of Mortágua et al. \cite{mortagua_enhancing_2024} with EAP-OAUTH, and the work of Marques et al. \cite{marques_eap-sh_2020} with EAP-SH. The former integrates web-based OAuth 2.0 authorisation to enable the user to access the network via EAP after web authentication in an Identity Provider. EAP-SH brings the captive portal web-based authentication to EAP, displaying a captive portal web page to authenticate the user, and then authorising the access to the network.

Both approaches rely on off-channel authentication based on the web, to later authorise the access in the EAP session. Our approach in EAP-FIDO is different, as the FIDO2 authentication messages are embedded in the EAP session, so the complete operation occurs during this connection. 

In this work, we have followed the methodology used in the work of Mortágua et al. \cite{mortagua_enhancing_2024} for the prototype implementation and performance analysis. Similar to their work, we implemented a prototype based on \texttt{hostap} well-known software codebase. Also, we conducted similar performance tests, so they can be compared. 

However, EAP-FIDO and EAP-OAUTH have a very different design. Their design of the protocol relies on off-channel OAuth 2.0 authentication, while ours on FIDO2. Despite this, we have adopted their strategy by designing EAP-FIDO as a tunnelled EAP method, and an adapted message structure, also using operation codes to keep the inner state. Also, we considered the idea of the ECDH key exchange for later key derivation, but binding the exchange to the FIDO2 authentication flow (see section~\ref{subsec:eap-fido:operation}).

\section{Conclusions and future work}
\label{sec:conclusions-and-future-work}

This paper presents our novel proposal: the EAP-FIDO method for Extensible Authentication Protocol (EAP) network authentication. This section concludes with the summary of the contributions.

\subsection{Our contribution: EAP-FIDO}
In this work, we have designed EAP-FIDO centred in compatibility. EAP-FIDO is based on the FIDO2 standards, so it is compatible with FIDO2 client and server libraries, as well as FIDO2 authenticators and credentials. EAP-FIDO is an EAP method, so its deployment in any 802.1X protected network authentication scenarios is direct, as EAP is designed to be extensible. Therefore, it does not require any implementation in Access Points (APs).

To showcase this design, we have implemented a prototype based on \texttt{hostap}, a software codebase implementing the complete EAP stack, including the Authentication Server (AS) and the Supplicant (client) \cite{malinen_hostapd_nodate}, where we have implemented EAP-FIDO. After the development, we measured its performance in a real Wi-Fi deployment.

Finally, we provided a security analysis and two possible use cases identified so far in corporate networks and public hotspots.

\subsection{Future work}
EAP-FIDO has been implemented as a prototype, but its functionality is limited. A more integrated and functional prototype could include a registration application through a web interface, and user management web interfaces for discoverable and non-discoverable credentials. Additionally, the prototype could be deployed to other operating systems, like Windows or Android, studying the availability of library dependencies for the FIDO2 client support.

Also as future work, we can further analyse the performance in real scenarios with several users, and study its usability among the users. This includes tests to assess the operation of the fast re-authentication mechanism or the usability of the authentication flow in daily usage.

{\small \section*{CRediT authorship contribution statement}
	\textbf{Martiño Rivera-Dourado}: Conceptualization, Investigation, Methodology, Project administration, Software, Writing – original draft, Writing review \& editing. \textbf{Christos Xenakis}: Investigation, Methodology, Resources, Supervision, Writing – review \& editing. \textbf{Alejandro Pazos}: Investigation, Methodology, Resources, Supervision, Funding acquisition. \textbf{Jose Vázquez-Naya}: Investigation, Methodology, Project administration, Resources, Supervision, Writing – review \& editing.}
	
{\small \section*{Declaration of competing interest}
The authors declare that they have no known competing financial interests or personal relationships that could have appeared to influence the work reported in this paper.}

{\small \section*{Acknowledgements}

This work was founded by EU and "Xunta de Galicia" (Spain), grant ED431C 2022/46–Competitive Reference Groups GRC. This work was also supported by CITIC, as a center accredited for excellence within the Galician University System and a member of the CIGUS Network, which receives subsidies from the Department of Education, Science, Universities, and Vocational Training of the "Xunta de Galicia". Additionally, CITIC is co-financed by the EU through the FEDER Galicia 2021-27 operational program (Ref. ED431G 2023/01). This research has been partially funded by the European Union's research and innovation programme under grant agreement No. 101095634 (ENTRUST). The work is also founded by the "Formación de Profesorado Universitario" (FPU) grant from the Spanish Ministry of Universities to Martiño Rivera Dourado (Grant FPU21/04519).}

\appendix
%



\bibliographystyle{elsarticle-num} 
\bibliography{main.bib}

\begin{thebibliography}{10}
\expandafter\ifx\csname url\endcsname\relax
  \def\url#1{\texttt{#1}}\fi
\expandafter\ifx\csname urlprefix\endcsname\relax\def\urlprefix{URL }\fi
\expandafter\ifx\csname href\endcsname\relax
  \def\href#1#2{#2} \def\path#1{#1}\fi

\bibitem{angelogianni_how_2024}
A.~Angelogianni, I.~Politis, C.~Xenakis,
  \href{https://dl.acm.org/doi/10.1145/3654661}{How many {FIDO} protocols are
  needed? analysing the technology, security and compliance}Just Accepted.
\newblock \href {https://doi.org/10.1145/3654661} {\path{doi:10.1145/3654661}}.
\newline\urlprefix\url{https://dl.acm.org/doi/10.1145/3654661}

\bibitem{lassak_why_2024}
L.~Lassak, E.~Pan, B.~Ur, M.~Golla,
  \href{https://www.usenix.org/conference/usenixsecurity18/presentation/bui}{Why
  aren't we using passkeys? obstacles companies face deploying {FIDO}2
  passwordless authentication}, in: 33th {USENIX} Security Symposium, {USENIX}
  Association.
\newline\urlprefix\url{https://www.usenix.org/conference/usenixsecurity18/presentation/bui}

\bibitem{veroni_large-scale_2022}
E.~Veroni, C.~Ntantogian, C.~Xenakis,
  \href{https://www.sciencedirect.com/science/article/pii/S2214212622000722}{A
  large-scale analysis of wi-fi passwords} 67  103190.
\newblock \href {https://doi.org/10.1016/j.jisa.2022.103190}
  {\path{doi:10.1016/j.jisa.2022.103190}}.
\newline\urlprefix\url{https://www.sciencedirect.com/science/article/pii/S2214212622000722}

\bibitem{wahab_investigating_2024}
F.~Wahab, I.~Khan, K.~Si,
  \href{https://doi.org/10.33436/v34i1y202408}{Investigating offline password
  attacks: A comprehensive review of rainbow table techniques and
  countermeasure limitations} 34~(1)  81--96.
\newblock \href {https://doi.org/10.33436/v34i1y202408}
  {\path{doi:10.33436/v34i1y202408}}.
\newline\urlprefix\url{https://doi.org/10.33436/v34i1y202408}

\bibitem{bradley_web_2021}
J.~Bradley, C.~Brand, A.~Langley, G.~Mandyam, N.~Satragno, N.~Steele, J.~Tan,
  S.~Weeden, M.~West, J.~Yasskin, \href{https://www.w3.org/TR/webauthn-2/}{Web
  authentication: An {API} for accessing public key credentials - level 2}.
\newline\urlprefix\url{https://www.w3.org/TR/webauthn-2/}

\bibitem{brand_web_2019}
C.~Brand, A.~Langley, G.~Mandyam, M.~West, J.~Yasskin,
  \href{https://www.w3.org/TR/webauthn-1/}{Web authentication: An {API} for
  accessing public key credentials - level 1}.
\newline\urlprefix\url{https://www.w3.org/TR/webauthn-1/}

\bibitem{rivera-dourado_novel_2024}
M.~Rivera-Dourado, M.~Gestal, A.~Pazos, J.~Vázquez-Naya,
  \href{https://www.mdpi.com/2076-3417/14/9/3610}{A novel protocol using
  captive portals for {FIDO}2 network authentication} 14~(9)  3610.
\newblock \href {https://doi.org/10.3390/app14093610}
  {\path{doi:10.3390/app14093610}}.
\newline\urlprefix\url{https://www.mdpi.com/2076-3417/14/9/3610}

\bibitem{huseynov_passwordless_2022}
E.~Huseynov, \href{https://ieeexplore.ieee.org/document/9984888/}{Passwordless
  {VPN} using {FIDO}2 security keys: Modern authentication security for legacy
  {VPN} systems}, in: 2022 4th International Conference on Data Intelligence
  and Security ({ICDIS}), {IEEE}, pp. 01--03, 3 citations (Semantic
  Scholar/{DOI}) [2024-02-15] 1 citations (Crossref) [2023-11-23].
\newblock \href {https://doi.org/10.1109/ICDIS55630.2022.00075}
  {\path{doi:10.1109/ICDIS55630.2022.00075}}.
\newline\urlprefix\url{https://ieeexplore.ieee.org/document/9984888/}

\bibitem{noauthor_ieee_2020}
\href{https://ieeexplore.ieee.org/document/9018454}{{IEEE} standard for local
  and metropolitan area networks–port-based network access control},
  conference Name: {IEEE} Std 802.1X-2020 (Revision of {IEEE} Std 802.1X-2010
  Incorporating {IEEE} Std 802.1Xbx-2014 and {IEEE} Std 802.1Xck-2018).
\newblock \href {https://doi.org/10.1109/IEEESTD.2020.9018454}
  {\path{doi:10.1109/IEEESTD.2020.9018454}}.
\newline\urlprefix\url{https://ieeexplore.ieee.org/document/9018454}

\bibitem{vollbrecht_extensible_2004}
J.~Vollbrecht, J.~D. Carlson, L.~Blunk, B.~D. Aboba, H.~Levkowetz,
  \href{https://datatracker.ietf.org/doc/rfc3748}{Extensible authentication
  protocol ({EAP})}.
\newblock \href {https://doi.org/10.17487/RFC3748}
  {\path{doi:10.17487/RFC3748}}.
\newline\urlprefix\url{https://datatracker.ietf.org/doc/rfc3748}

\bibitem{mortagua_enhancing_2024}
D.~Mortágua, A.~Zúquete, P.~Salvador,
  \href{https://linkinghub.elsevier.com/retrieve/pii/S1389128624001695}{Enhancing
  802.1x authentication with identity providers using {EAP}-{OAUTH} and {OAuth}
  2.0} 244  110337.
\newblock \href {https://doi.org/10.1016/j.comnet.2024.110337}
  {\path{doi:10.1016/j.comnet.2024.110337}}.
\newline\urlprefix\url{https://linkinghub.elsevier.com/retrieve/pii/S1389128624001695}

\bibitem{armstrong_client_2022}
C.~Armstrong, K.~Georgantas, F.~Kaczmarczyck, N.~Satragno, N.~Sung,
  \href{https://fidoalliance.org/specs/fido-v2.1-ps-20210615/fido-client-to-authenticator-protocol-v2.1-ps-errata-20220621.html}{Client
  to authenticator protocol ({CTAP})}.
\newline\urlprefix\url{https://fidoalliance.org/specs/fido-v2.1-ps-20210615/fido-client-to-authenticator-protocol-v2.1-ps-errata-20220621.html}

\bibitem{calhoun_radius_2003}
P.~R. Calhoun, B.~D. Aboba,
  \href{https://datatracker.ietf.org/doc/rfc3579}{{RADIUS} (remote
  authentication dial in user service) support for extensible authentication
  protocol ({EAP})}.
\newblock \href {https://doi.org/10.17487/RFC3579}
  {\path{doi:10.17487/RFC3579}}.
\newline\urlprefix\url{https://datatracker.ietf.org/doc/rfc3579}

\bibitem{funk_extensible_2008}
P.~Funk, S.~Blake-Wilson,
  \href{https://datatracker.ietf.org/doc/rfc5281}{Extensible authentication
  protocol tunneled transport layer security authenticated protocol version 0
  ({EAP}-{TTLSv}0)}.
\newblock \href {https://doi.org/10.17487/RFC5281}
  {\path{doi:10.17487/RFC5281}}.
\newline\urlprefix\url{https://datatracker.ietf.org/doc/rfc5281}

\bibitem{palekar_protected_2004}
A.~Palekar, S.~Josefsson, D.~Simon, G.~Zorn,
  \href{https://datatracker.ietf.org/doc/draft-josefsson-pppext-eap-tls-eap-10}{Protected
  {EAP} protocol ({PEAP}) version 2}, num Pages: 87.
\newline\urlprefix\url{https://datatracker.ietf.org/doc/draft-josefsson-pppext-eap-tls-eap-10}

\bibitem{simon_extensible_2008}
D.~Simon, B.~D. Aboba, P.~Eronen,
  \href{https://datatracker.ietf.org/doc/rfc5247}{Extensible authentication
  protocol ({EAP}) key management framework}.
\newblock \href {https://doi.org/10.17487/RFC5247}
  {\path{doi:10.17487/RFC5247}}.
\newline\urlprefix\url{https://datatracker.ietf.org/doc/rfc5247}

\bibitem{dekok_extensible_2020}
A.~{DeKok}, \href{https://datatracker.ietf.org/doc/rfc8940}{Extensible
  authentication protocol ({EAP}) session-id derivation for {EAP} subscriber
  identity module ({EAP}-{SIM}), {EAP} authentication and key agreement
  ({EAP}-{AKA}), and protected {EAP} ({PEAP})}.
\newblock \href {https://doi.org/10.17487/RFC8940}
  {\path{doi:10.17487/RFC8940}}.
\newline\urlprefix\url{https://datatracker.ietf.org/doc/rfc8940}

\bibitem{malinen_hostapd_nodate}
J.~Malinen, \href{https://w1.fi/hostapd/}{hostapd: {IEEE} 802.11 {AP}, {IEEE}
  802.1x/{WPA}/{WPA}2/{WPA}3/{EAP}/{RADIUS} authenticator}, w1f1.
\newline\urlprefix\url{https://w1.fi/hostapd/}

\bibitem{martelletto_libfido2_nodate}
P.~Martelletto, L.~Michaelsson,
  \href{https://developers.yubico.com/libfido2/}{libfido2}.
\newline\urlprefix\url{https://developers.yubico.com/libfido2/}

\bibitem{chen_recommendations_2023}
L.~Chen, D.~Moody, A.~Regenscheid, A.~Robinson, K.~Randall,
  \href{https://nvlpubs.nist.gov/nistpubs/SpecialPublications/NIST.SP.800-186.pdf}{Recommendations
  for discrete logarithm-based cryptography: Elliptic curve domain parameters}.
\newblock \href {https://doi.org/10.6028/NIST.SP.800-186}
  {\path{doi:10.6028/NIST.SP.800-186}}.
\newline\urlprefix\url{https://nvlpubs.nist.gov/nistpubs/SpecialPublications/NIST.SP.800-186.pdf}

\bibitem{palama_attacks_2023}
I.~Palamà, A.~Amici, G.~Bellicini, F.~Gringoli, F.~Pedretti, G.~Bianchi,
  \href{https://www.sciencedirect.com/science/article/pii/S014036642300347X}{Attacks
  and vulnerabilities of wi-fi enterprise networks: User security awareness
  assessment through credential stealing attack experiments} 212  129--140.
\newblock \href {https://doi.org/10.1016/j.comcom.2023.09.031}
  {\path{doi:10.1016/j.comcom.2023.09.031}}.
\newline\urlprefix\url{https://www.sciencedirect.com/science/article/pii/S014036642300347X}

\bibitem{malkin_provable_2021}
M.~Barbosa, A.~Boldyreva, S.~Chen, B.~Warinschi,
  \href{https://link.springer.com/10.1007/978-3-030-84252-9_5}{Provable
  security analysis of {FIDO}2}, in: T.~Malkin, C.~Peikert (Eds.), Advances in
  Cryptology – {CRYPTO} 2021, Vol. 12827, Springer International Publishing,
  pp. 125--156, 34 citations (Google Scholar).
\newline\urlprefix\url{https://link.springer.com/10.1007/978-3-030-84252-9_5}

\bibitem{silhavy_fido2_2023}
S.~A. Dixit, A.~Gupta, R.~Jain, R.~Joshi, S.~Gonge, K.~Kotecha,
  \href{https://link.springer.com/10.1007/978-3-031-35317-8_32}{{FIDO}2
  passwordless authentication for remote devices}, in: R.~Silhavy, P.~Silhavy
  (Eds.), Networks and Systems in Cybernetics, Vol. 723, Springer International
  Publishing, pp. 349--362, 0 citations (Google Scholar).
\newline\urlprefix\url{https://link.springer.com/10.1007/978-3-031-35317-8_32}

\bibitem{luo_g2f_2021}
H.~Luo, C.~Wang, H.~Luo, F.~Zhang, F.~Lin, G.~Xu,
  \href{https://ieeexplore.ieee.org/document/9319261/}{G2f: A secure user
  authentication for rapid smart home {IoT} management} 8~(13)  10884--10895,
  16 citations (Semantic Scholar/{DOI}) [2024-02-15] 13 citations (Crossref)
  [2023-11-23].
\newblock \href {https://doi.org/10.1109/JIOT.2021.3050710}
  {\path{doi:10.1109/JIOT.2021.3050710}}.
\newline\urlprefix\url{https://ieeexplore.ieee.org/document/9319261/}

\bibitem{chakkaravarthy_sethuraman_loki_2022}
S.~Chakkaravarthy~Sethuraman, A.~Mitra, K.-C. Li, A.~Ghosh, M.~Gopinath,
  N.~Sukhija, \href{https://ieeexplore.ieee.org/document/9927414/}{Loki: A
  physical security key compatible {IoT} based lock for protecting physical
  assets} 10  112721--112730, 1 citations (Semantic Scholar/{DOI}) [2024-02-15]
  1 citations (Crossref) [2023-11-23].
\newblock \href {https://doi.org/10.1109/ACCESS.2022.3216665}
  {\path{doi:10.1109/ACCESS.2022.3216665}}.
\newline\urlprefix\url{https://ieeexplore.ieee.org/document/9927414/}

\bibitem{dammak_secure_2020}
M.~Dammak, S.~Aroua, S.~M. Senouci, Y.~Ghamri-Doudane, G.~Suciu, M.-A. Sachian,
  R.~Roscaneanu, I.~Ozkan, M.~O. Gungor,
  \href{https://ieeexplore.ieee.org/document/9248483/}{A secure and
  interoperable platform for privacy protection in the smart hotel context},
  in: 2020 Global Information Infrastructure and Networking Symposium ({GIIS}),
  {IEEE}, pp. 1--6, 2 citations (Semantic Scholar/{DOI}) [2024-02-15] 1
  citations (Crossref) [2023-11-23].
\newblock \href {https://doi.org/10.1109/GIIS50753.2020.9248483}
  {\path{doi:10.1109/GIIS50753.2020.9248483}}.
\newline\urlprefix\url{https://ieeexplore.ieee.org/document/9248483/}

\bibitem{conners_lets_2022}
J.~Conners, C.~Devenport, S.~Derbidge, N.~Farnsworth, K.~Gates, S.~Lambert,
  C.~{McClain}, P.~Nichols, D.~Zappala,
  \href{https://www.ndss-symposium.org/wp-content/uploads/2022-272-paper.pdf}{Let’s
  authenticate: Automated certificates for user authentication}, in:
  Proceedings 2022 Network and Distributed System Security Symposium, Internet
  Society, 5 citations (Semantic Scholar/{DOI}) [2024-02-15].
\newblock \href {https://doi.org/10.14722/ndss.2022.24272}
  {\path{doi:10.14722/ndss.2022.24272}}.
\newline\urlprefix\url{https://www.ndss-symposium.org/wp-content/uploads/2022-272-paper.pdf}

\bibitem{chifor_flexible_2018}
B.-C. Chifor, S.~Teican, M.~Togan, G.~Gugulea,
  \href{https://ieeexplore.ieee.org/document/8484268/}{A flexible authorization
  mechanism for enterprise networks using smart-phone devices}, in: 2018
  International Conference on Communications ({COMM}), {IEEE}, pp. 437--440, 4
  citations (Semantic Scholar/{DOI}) [2024-02-15] 1 citations (Crossref)
  [2023-11-23].
\newblock \href {https://doi.org/10.1109/ICComm.2018.8484268}
  {\path{doi:10.1109/ICComm.2018.8484268}}.
\newline\urlprefix\url{https://ieeexplore.ieee.org/document/8484268/}

\bibitem{panizza_fido2_2024}
J.~Panizza,
  \href{http://sar.informatik.hu-berlin.de/research/publications/SAR-PR-2024-02/SAR-PR-2024-02_.pdf}{{FIDO}2
  {TLS} 1.3 extension: Strong {EAP}-{TLS} authentication for 802.1x networks}.
\newline\urlprefix\url{http://sar.informatik.hu-berlin.de/research/publications/SAR-PR-2024-02/SAR-PR-2024-02_.pdf}

\bibitem{rieckers_eap-fido_nodate}
J.-F. Rieckers, S.~Winter,
  \href{https://datatracker.ietf.org/doc/draft-ietf-emu-eap-fido/}{{EAP}-{FIDO}},
  num Pages: 28.
\newline\urlprefix\url{https://datatracker.ietf.org/doc/draft-ietf-emu-eap-fido/}

\bibitem{vasileios_grammatopoulos_web_2021}
A.~Vasileios~Grammatopoulos, I.~Politis, C.~Xenakis,
  \href{https://dl.acm.org/doi/10.1145/3465481.3469209}{A web tool for
  analyzing {FIDO}2/{WebAuthn} requests and responses}, in: Proceedings of the
  16th International Conference on Availability, Reliability and Security,
  {ACM}, pp. 1--10.
\newblock \href {https://doi.org/10.1145/3465481.3469209}
  {\path{doi:10.1145/3465481.3469209}}.
\newline\urlprefix\url{https://dl.acm.org/doi/10.1145/3465481.3469209}

\bibitem{grammatopoulos_blind_2022}
A.~V. Grammatopoulos, I.~Politis, C.~Xenakis,
  \href{https://isyou.info/jowua/papers/jowua-v13n2-4.pdf}{Blind
  software-assisted conformance and security assessment of {FIDO}2/{WebAuthn}
  implementations.} 13~(2)  96--127.
\newline\urlprefix\url{https://isyou.info/jowua/papers/jowua-v13n2-4.pdf}

\bibitem{marques_eap-sh_2020}
N.~Marques, A.~Zúquete, J.~P. Barraca,
  \href{https://doi.org/10.1007/s11277-020-07298-y}{{EAP}-{SH}: An {EAP}
  authentication protocol to integrate captive portals in the 802.1x security
  architecture} 113~(4)  1891--1915, number: 4.
\newblock \href {https://doi.org/10.1007/s11277-020-07298-y}
  {\path{doi:10.1007/s11277-020-07298-y}}.
\newline\urlprefix\url{https://doi.org/10.1007/s11277-020-07298-y}

\end{thebibliography}






\end{document}